\documentclass[fleqn,usenatbib]{mnras}

\usepackage{newtxtext,newtxmath}
\usepackage{ulem}

\usepackage[T1]{fontenc}

\usepackage{graphicx}	
\usepackage{amsmath}	
\usepackage{xcolor}
\usepackage{array}

\definecolor{pgreen}{RGB}{46, 121, 10}
\setcitestyle{notesep={}}
\newcommand{\shark}{\textsc{shark}}
\newcommand{\sharkfit}{\textsc{shark$_\mathrm{fit}$}}

\newcommand{\prospect}{\textsc{ProSpect}}

\newcommand{\comment}[1]{\textbf{\textcolor{pgreen}{Mat\'ias: #1}}}

\newcommand{\mstar}[1]{$10^{#1}$ M$_\odot$}
\newcommand{\PR}{$P^{}_\mathrm{R}$}
\newcommand{\tLBB}{$t^{}_\mathrm{LB,B}$}
\newcommand{\tLBR}{$t^{}_\mathrm{LB,R}$}
\newcommand{\tauQ}{$\tau^{}_\mathrm{Q}$}

\defcitealias{bravo2022}{Paper I}

\title[Forensic quenching timescales]{Galaxy quenching timescales from a forensic reconstruction of their colour evolution}

\author[Bravo et al.]{
Mat\'ias Bravo$^{1,2}$\thanks{E-mail:bravosam@mcmaster.ca},
Aaron S.~G. Robotham$^{1,3}$,
Claudia del P. Lagos$^{1,3}$,
\newauthor{
Luke J. M. Davies$^1$,
Sabine Bellstedt$^1$ and
Jessica E. Thorne$^1$}
\\
$^{1}$International Centre for Radio Astronomy Research (ICRAR), M468, University of Western Australia, 35 Stirling Hwy, Crawley, \\WA 6009, Australia.\\
$^{2}$Department of Physics \& Astronomy, McMaster University, 1280 Main Street W, Hamilton, ON, L8S 4M1, Canada\\
$^{3}$ARC Centre of Excellence for All Sky Astrophysics in 3 Dimensions (ASTRO 3D).\\
}

\date{Accepted XXX. Received YYY; in original form ZZZ}

\pubyear{2023}

\begin{document}
\label{firstpage}
\pagerange{\pageref{firstpage}--\pageref{lastpage}}
\maketitle

\begin{abstract}
The timescales on which galaxies move out of the blue cloud to the red sequence (\tauQ) provide insight into the mechanisms driving quenching.
Here, we build upon previous work, where we showcased a method to reconstruct the colour evolution of observed low-redshift galaxies from the Galaxy And Mass Assembly (GAMA) survey based on spectral energy distribution (SED) fitting with \prospect, together with a statistically-driven definition for the blue and red populations.
We also use the predicted colour evolution from the \shark\ semi-analytic model, combined with SED fits of our simulated galaxy sample, to study the accuracy of the measured \tauQ\ and gain physical insight into the colour evolution of galaxies.
In this work, we measure \tauQ\ in a consistent approach for both observations and simulations.
After accounting for selection bias, we find evidence for an increase in \tauQ\ in GAMA as a function of cosmic time (from \tauQ$\sim1$ Gyr to \tauQ$\sim2$ Gyr in the lapse of $\sim4$ Gyr), but not in \shark\ (\tauQ$\lesssim1$ Gyr).
Our observations and simulations disagree on the effect of stellar mass, with GAMA showing massive galaxies transitioning faster, but is the opposite in \shark.
We find that environment only impacts galaxies below $\sim$\mstar{10} in GAMA, with satellites having shorter \tauQ\ than centrals by $\sim0.4$ Gyr, with \shark\ only in qualitative agreement.
Finally, we compare to previous literature, finding consistency with timescales in the order of couple Gyr, but with several differences that we discuss.
\end{abstract}


\section{Introduction}\label{sec:intro}

One of the most striking features of galaxies in the local Universe is the optical colour bimodality \citep[e.g.,][]{strateva2001,blanton2003,baldry2004,driver2006}, with most galaxies being either blue or red.
Compared to these populations, there are comparatively few galaxies in the intermediate region, often referred to as the "green valley", \citep[e.g.,][]{martin2007,wyder2007,schawinski2014}.
Stars are the dominant source of the light emitted by most galaxies (at low redshift), suggesting that this bimodality is a consequence of the presence of two dominant stellar populations for galaxies.
As the (intrinsic) colour of stars is mainly driven by their age, the colour bimodality is a reflection of a bimodality in the recent star formation in galaxies.

These populations are also characterised by intrinsically different galaxy properties.
Red galaxies are preferentially of early-type morphology \citep[e.g.,][]{bershady2000,mignoli2009,schawinski2014}, more massive \citep[e.g.,][]{baldry2004,peng2010,taylor2015}, and found in denser environments \citep[e.g.,][]{kauffmann2004,baldry2006,peng2010}.
Studies have also shown that this bimodality is seen across cosmic time, with the fraction of galaxies in the red population increasing towards recent times \citep[e.g.,][]{wolf2003,bell2004,williams2009}.
It has also been found that the first galaxies that joined the red population are more massive than those that have joined at more recent times, a process called downsizing \citep[e.g.,][]{cowie1996,brinchmann2000,heavens2004}.
Combined, these observations present a broad picture where galaxies grow as part of the star-forming blue population, with some of them eventually ceasing to form stars and joining the red population \citep[commonly referred to as quenching, e.g.,][]{bell2004,blanton2006,faber2007}.

The relative lack of galaxies located in the green valley implies short timescales to transition in colour (quench) for the galaxies that join the red population \citep[e.g.,][]{schawinski2014,bremer2018}.
Different mechanisms to quench star formation are expected to do so on different timescales \citep[e.g.,][]{kaviraj2011,wetzel2013,schawinski2014,wheeler2014}, hence, studying these timescales can offer a view into the physical processes that govern galaxy evolution.
Theoretical models are a critical tool to explore these mechanism, as we gain insight by testing their predictions against results from observations.
A well-known example in the literature is that a quenching mechanism capable of stopping gas accretion onto galaxies is required to produce massive red galaxies, usually assumed to be driven by active galactic nuclei or shock heating of the halo gas \citep[e.g.,][]{bower2006,cattaneo2006,croton2006,lagos2008}.

A historical challenge for simulations has been their inability to produce colour distributions well-matched to observations \citep[e.g.,][]{weinmann2006,font2008,coil2008}, though recent advances have largely ameliorated this tensions \citep[e.g.,][]{trayford2015,nelson2018a,lagos2019,bravo2020}.
These advances now enable the exploration of the colour evolution of galaxies with theoretical models, leading to the prediction of the timescales on which galaxies transition from being blue to red \citep[e.g.,][]{trayford2016,nelson2018a,wright2019}.
This colour evolution is not directly measurable from observations and can only be inferred \citep[e.g.,][]{schawinski2014,smethurst2015,rowlands2018,phillipps2019}.
This means that results cannot be directly compared to the predictions of theoretical models.
Further complicating comparisons is the lack of a unified definition for how to measure the colour transition timescales, or even what galaxies should be classified as blue or red \citep[e.g., see classifications by][]{martin2007,schawinski2014,taylor2015,bremer2018,wright2019}.

In \citet[, hereafter Paper I]{bravo2022}, we described a novel method to reconstruct the colour evolution of low-redshift observed galaxies from the Galaxy And Mass Assembly \citep[GAMA;][]{driver2011,liske2015} survey, using the \prospect\ spectral energy distribution (SED) fitting tool \citep{robotham2020}.
We tested this recovery by performing the same procedure with a comparable sample of galaxies generated with the \shark\ semi-analytic model \citep[SAM;][]{lagos2018,lagos2019}, finding that we can accurately recover the colour evolution of the last $\sim6$ Gyr for galaxies with current masses above $\sim$\mstar{9}.
Finally, we provided a statistically-motivated definition for the blue and red populations, their evolution through cosmic time, and demonstrated the resulting probabilities of galaxies belonging to either blue or red population.

In this work we now utilise these results to explore how quickly galaxies transition from being blue to red, in a novel approach that is consistent and directly comparable for both observations and simulations.
In Section \ref{sec:probabilities} we explore the distribution of probabilities of galaxies being red, to construct statistically-motivated definitions for when a galaxy is certainly a member of of either population, and when is transitioning between both.
We then use that classification to explore the timescale on which galaxies transitioned from blue to red in Section \ref{sec:timescales}, exploring possible time, mass, and environmental effects.
For conciseness, we will refer to this blue-to-red transition timescale as \tauQ\ throughout this work.
In Section \ref{sec:disc} we discuss our results, both for the physical implications of the timescales we measure and to compare with the existing literature.
Finally, we present our conclusions in Section \ref{sec:summary}.
In this work, we adopt the \citet{planck2016xiii} $\Lambda$CDM cosmology, with values of matter, baryon, and dark energy densities of $\Omega_b=0.0488$, and $\Omega_\Lambda=0.6879$, respectively, and a Hubble parameter of H$_0=67.51$ km s$^{-1}$ Mpc$^{-1}$.

\section{Galaxy catalogues}\label{sec:GSP}

\begin{figure*}
    \centering
    \includegraphics[width=\linewidth]{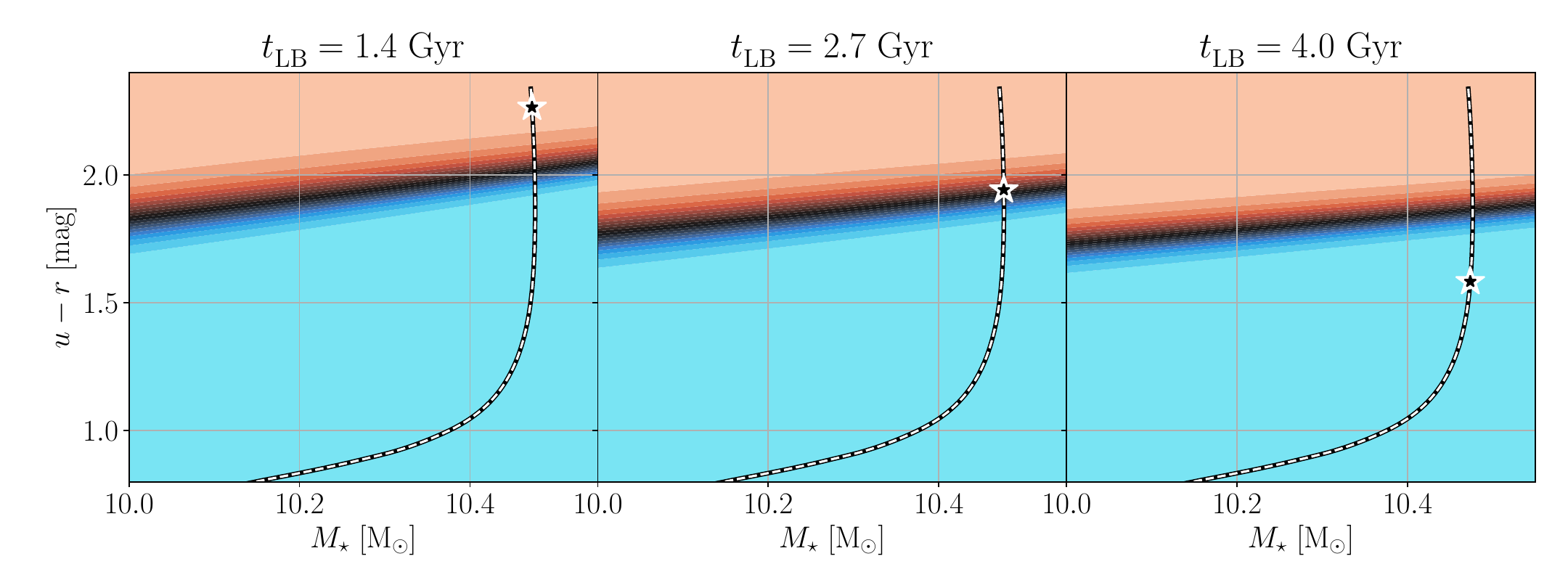}
    \caption{The colour evolution of a single galaxy and its transition from the blue to the red population from the results  \citetalias{bravo2022}, with the galaxy \texttt{CATAID=92739} from the GAMA survey used for this example.
    Each panels shows a small section of the colour-mass plane for the GAMA survey at three different lookback times, with the coloured contours showing the probability of being red for a a galaxy in any given position in this space. 
    The complete evolution track of the galaxy is shown by the black and white line running from the bottom left to the top right of each panel, with the position of the galaxy in the corresponding lookback time of each panel shown with a star marker.}
    \label{fig:intro_fig}
\end{figure*}

In this work, we use the data set presented in \citetalias{bravo2022}, which we briefly outline.
This data set is composed of the intrinsic colour (i.e., not attenuated by dust) and stellar mass histories for three low-redshift galaxy samples, both derived from their star formation and metallicity histories (SFH and $Z$H, respectively).
The first one is comprised of $\sim7,000$ galaxies from the GAMA survey used in \citet{bellstedt2020b,bellstedt2021}.
The other two are each comprised of $\sim30,000$ GAMA-like galaxies (i.e., $r^{}\mathrm{apparent}<19.8$ mag) from the \shark\ SAM \citep{lagos2018,lagos2019}.
For the GAMA sample, we reconstructed their colour evolution from the star formation and metallicity histories inferred from the SED fitting by \citet{bellstedt2020b}\footnote{The SFH model adopted by \citet{bellstedt2020b} is a skewed Gaussian, a parametric model but significantly more flexible than other common parametric models in the literature \citep[e.g.,][]{dacunha2008,noll2009,carnall2018,boquien2019}, but still unable to model rejuvenation episodes.
    While not unique among SED fitting models \citep[e.g.,][]{carnall2018,johnson2021}, the use of \prospect\ in the literature has been unique in the assumption that gas metallicities \textit{evolves}, modelling it as a linear scaling of the mass growth of galaxies \citep[][]{bellstedt2020b,thorne2021,thorne2022}.}
by combining these histories with the stellar population synthesis model used for the fitting \citep{bruzual2003}.
The two \shark\ samples contain the same galaxies, the difference is how we constructed the colour and stellar mass evolution: one sample is the predicted evolution from the simulation itself (which we will refer as \shark); the other presents the inferred evolution from SED fitting the \shark\ galaxies with the same method as with the GAMA sample (we will refer to this sample as \sharkfit).
Section 2 of \citetalias{bravo2022} contains the detailed description of these three samples, with Appendix A offering a deeper exploration of our SED modelling choices for the interested reader.

Inspired by \citet{taylor2015}, in \citetalias{bravo2022} we modelled the colour-mass distribution for each sample with a time-and-mass-dependent Gaussian Mixture Model (GMM).
Consistent with the modelling by \citet{baldry2004} and \citet{taylor2015}, we described the colour-mass distribution of galaxies in \citetalias{bravo2022} with two evolving populations: blue and red\footnote{We did test using three components, but we found no statistical evidence for a third (green) population.
    See section 3.3 of \citetalias{bravo2022} for further details.}
.
These GMMs are described by five parameters: the relative fraction of blue (or red) galaxies, and the means and standard deviations of each population.
In \citetalias{bravo2022} we presented a two-step parameterisation of these parameters, first as a function of stellar mass, and second as a function of lookback time.
For the stellar mass parameterisation, based on the distributions of the GMM parameters as a function of stellar mass, we chose to parameterise the relative blue/red fractions with a logistic curve, and with first-order polynomials for the means and standard deviations.
We then parameterised the time evolution of the stellar mass-dependent Gaussian parameters, using second- and third-order polynomials.
Section 3 of \citetalias{bravo2022} provides the complete description of this modelling, with section 3.1, figure 2 and table 1 offering an simple overview.

With a complete parameterisation of the evolution of the colour populations for all three samples, we can calculate the probability for any galaxy belonging to the blue or red population at any given time.
As in figure 12 of \citetalias{bravo2022}, in this work we choose to show the probability of being red, \PR\footnote{Formally the probability of being red is a function of lookback time, stellar mass, and colour, but for brevity we will refer to it as \PR\ instead of \PR$(t^{}_\mathrm{LB},M_\star,u-r$).}
, which is calculated from the Gaussian Mixture Model with which we model the galaxy colour distribution (see sections 3.3 and 3.4 of \citetalias{bravo2022} for further details).

We also showed that our colour-based classification leads to a sensible separation in specific star formation rate as a function of stellar mass.
The transition zone is not cleanly defined in this space, as expected from the scatter between colour and specific star formation rate.
In \citetalias{bravo2022}, through the comparison of the colour evolution of the three samples (GAMA, \shark, and \sharkfit), we found that the reconstruction of the colour evolution becomes biased by the modelling choices in the SED fitting for lookback times above $\gtrsim6$ Gyr.
For this reason, while we will measure colour evolution from a lookback time of 10 Gyr onward and show some of our results at higher lookback times, we mainly focus on the \tauQ\ we measure below a lookback time of 6 Gyr.

In this work, we use \PR\ to calculate \tauQ, defining the lookback times when a galaxy leaves the blue population (\tLBB) and when it joins the red population (\tLBR), which are related to \tauQ\ as:
\begin{equation}
    \tau^{}_\mathrm{Q}=t^{}_\mathrm{LB,B}-t^{}_\mathrm{LB,R},\label{eq:taudef}
\end{equation}
\noindent where we define these lookback times such that $t^{}_\mathrm{LB,B}>t^{}_\mathrm{LB,R}$ (i.e., \tauQ$>0$).
In \citetalias{bravo2022} we chose a time step of 100 Myr to reconstruct the colour evolution of galaxies, meaning that the shortest measurable \tauQ\ is 100 Myr.
Figure \ref{fig:intro_fig} shows an example of the data set we constructed in \citetalias{bravo2022} and use in this work to measure \tauQ, with both the evolutionary tracks in the colour-mass plane of individual galaxies and our model to calculate \PR\ at any point in the $t^{}_\mathrm{LB}$--$M_\star$--$(u-r)$ space.
Our example galaxy moves from being likely blue until a lookback time of $\sim4$, to have a similar probability of being either blue or red at $\sim2.7$ Gyr (\PR$\sim0.5$), to likely becoming red at $\sim1.5$ Gyr.
What is not immediately obvious from this Figure alone is what values of \PR\ best define \tLBB\ and \tLBR, which is the first aspect we will address in Section 3.

\section{Distribution and evolution of the probability of galaxies being red}\label{sec:probabilities}

\begin{figure}
    \centering
    \includegraphics[width=\linewidth]{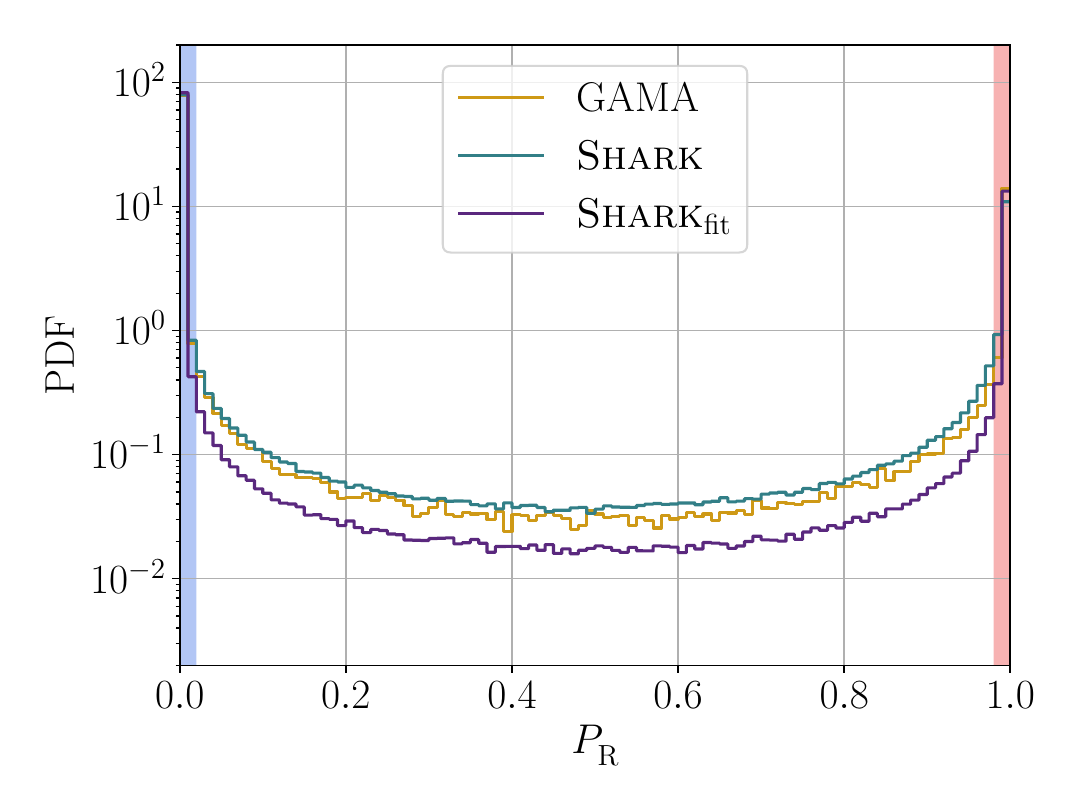}
    \caption{Distribution of probability of being red for all galaxies and all time steps below 6 Gyr.
    As in \citetalias{bravo2022}, the orange line shows the distribution for GAMA, cyan for \shark, and purple for \sharkfit.
    Each bin spans 1\% in probability.
    Highlighted in blue/red are the probability ranges where we define a galaxy as being blue/red (\PR$=0.02$ and \PR$=0.98$, respectively).}
    \label{fig:PRdist}
\end{figure}

\begin{figure*}
    \centering
    \includegraphics[width=\linewidth]{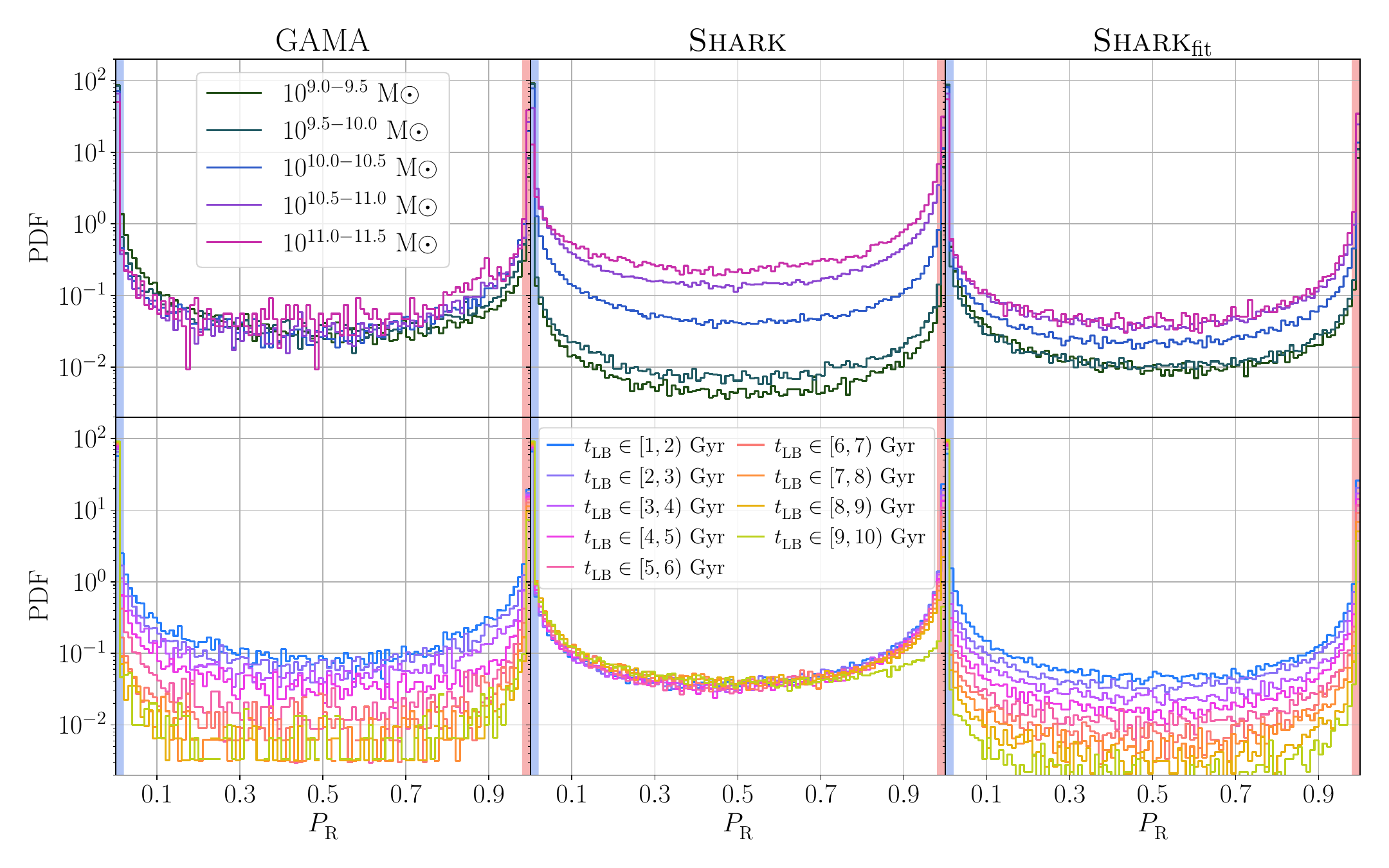}
    \caption{Distribution of the probability of galaxies being red. 
    Each column shows the probability distribution for single sample, from left to right: GAMA, \shark, and \sharkfit.
    The top row shows the distribution at all time steps below 6 Gyr binned by stellar mass at observation time, with bins of increasing mass shown with lighter colours.
    The bottom row shows the distribution at all stellar masses binned by lookback time, with bins of increasing lookback time in lighter colours.
    Highlighted in blue (red) are the probability ranges where we define a galaxy as being blue (red), as in Figure \ref{fig:PRdist}.}
    \label{fig:PRdist_bymass&time}
\end{figure*}

With the probabilistic blue/red classification from \citetalias{bravo2022}, the last step needed to measure \tauQ\ is the choice of probabilities at which a galaxy is considered to be a part of the blue or red populations.
While ultimately this is an arbitrary choice, we will use the distribution of our calculated probabilities to inform this choice, just like our choice of a GMM to describe the colour populations in \citetalias{bravo2022} was informed by the reconstructed colour distributions.
As the green valley is sparsely populated, most of the mass of the PDF will be near the edges (i.e., near \PR$=0$ and \PR$=1$).
We use the second derivative of the decrease of the PDF from the edges towards the centre as a guide for our choice.
Figure \ref{fig:PRdist} shows the distribution of probabilities of being red (\PR), stacked from all time steps and stellar masses given our selection criteria.
The transition from the extremes of the probability range is dramatic, with a $\sim2$ dex ($\sim1$ dex) decrease in the PDF from the 0-1\% bin (99-100\%) to the 1-2\% bin (98-99\%).
This would suggest \PR$<0.01$ (\PR$>0.99$) is a reasonable criterion for a galaxy to be confidently classified as blue (red).

While such a selection will work on average, the distribution of probabilities may depend on stellar mass and/or lookback time.
To examine this, Figure \ref{fig:PRdist_bymass&time} shows the probability distributions for all samples binned by both stellar mass and lookback time.
GAMA and \shark\ show opposite trends, with the former exhibiting a probability distribution that is mass-independent but time-dependent, and the latter being mass-dependent but time-independent.
This difference suggest that we expect a stellar mass trend for \tauQ\ in \shark, and a lookback time trend in GAMA.
\sharkfit\ exhibits a mix of the trends in GAMA and \shark, suggesting that our modelling choices in \prospect\ may be impacting our \tauQ\ measurements, but also that they are not completely dictated by them.

While for most of the PDFs shown, a choice of \PR$<0.01$ (\PR$>0.99$) would still lead to a strong blue (red) classification, this is not true for the two highest mass bins in \shark.
For this reason we will use a more conservative classification of \PR$<0.02$ for blue galaxies, \PR$>0.98$ for red galaxies, for the rest of this work (see also the bottom row of Table \ref{tab:lit}).
We note that small variations to these limits do not affect the qualitative nature of our results, nor do they lead to strong quantitative changes.

\begin{figure*}
    \centering
    \includegraphics[width=\linewidth]{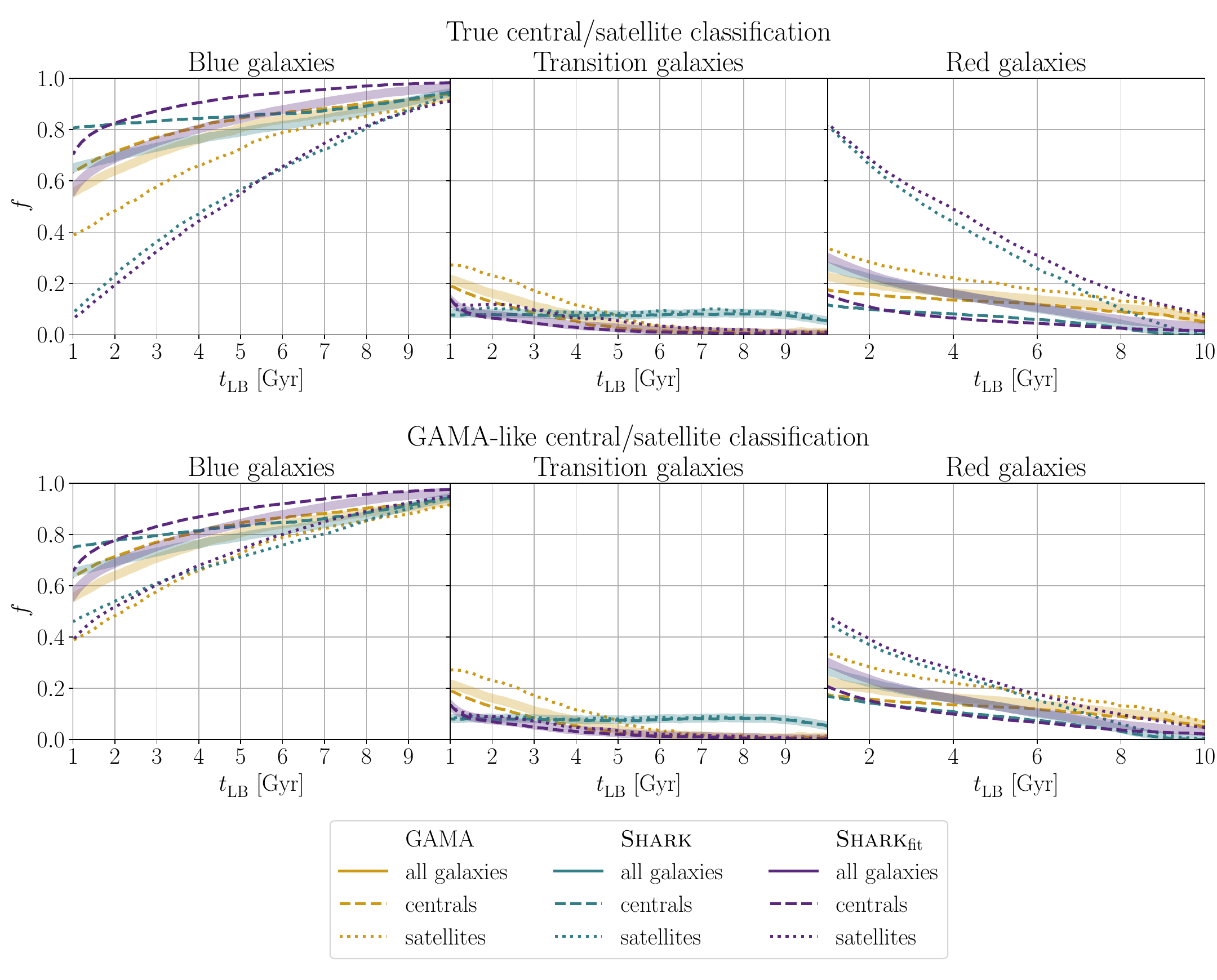}
    \caption{Time evolution of the fraction of galaxies classified as blue/transitional/red, as a function of the central/satellite classification for \shark\ and \sharkfit.
    Galaxies included correspond to those above the evolving mass completeness limits at $z\sim0.06$ defined in \citetalias{bravo2022} (see their section 3.2), corresponding to $M_\star(z\sim0.06)\geq$\mstar{9.1} for GAMA and $M_\star(z\sim0.06)\geq$\mstar{9.0} for \shark/\sharkfit.
    Each column shows a different population: blue in the left column, transitional in the middle, and red in the right.
    In each panel the corresponding population is shown for our three samples, with the combined central+satellite fraction is shown in solid lines, centrals only with dashed lines, and satellites with dotted lines.
    \shark\ and \sharkfit\ are shown in the top row using the central/satellite classification from the simulation, and using a GAMA-like classification in the bottom row \citep[$23\%$ confusion, following the results from][]{bravo2020,chauhan2021}.
    Line colours are as in Figure \ref{fig:PRdist}.
    Columns are as in Figure \ref{fig:PRdist_bymass&time}.
    The results for GAMA are identical in each column, they are repeated for easier comparison with both classifications from the simulations.}
    \label{fig:popfrac_censat}
\end{figure*}

\subsection{Time evolution of the fractions of blue, transitional, and red galaxies}\label{subsec:time_evol}

\begin{table*}
    \centering
    \begin{tabular}{c|c|c|c}
        Selected galaxies & GAMA & \textsc{Shark} & \textsc{Shark}$_\mathrm{fit}$ \\
        \hline
        Are red by $t^{}_\mathrm{LB}=1$ Gyr & 22.9\% & 26.7\% & 30.2\% \\
        Became red at $t^{}_\mathrm{LB}<10$ Gyr & 15.0\% & $<26.0\%$ & 21.2\% \\
        Were blue at $t^{}_\mathrm{LB}=10$ Gyr and are red at $t^{}_\mathrm{LB}=1$ Gyr & 14.3\% & 22.2\% & 21.0\% \\
    \end{tabular}
    \caption{Percentage of the total population of galaxies that are currently red, that became red after $t^{}_\mathrm{LB}=10$ Gyr, and that transitioned from blue to red after $t^{}_\mathrm{LB}=10$ Gyr, from all three samples.
    The bottom row is the sample selected to measure \tauQ.}
    \label{tab:prob_dist}
\end{table*}

\begin{figure*}
    \centering
    \includegraphics[width=\linewidth]{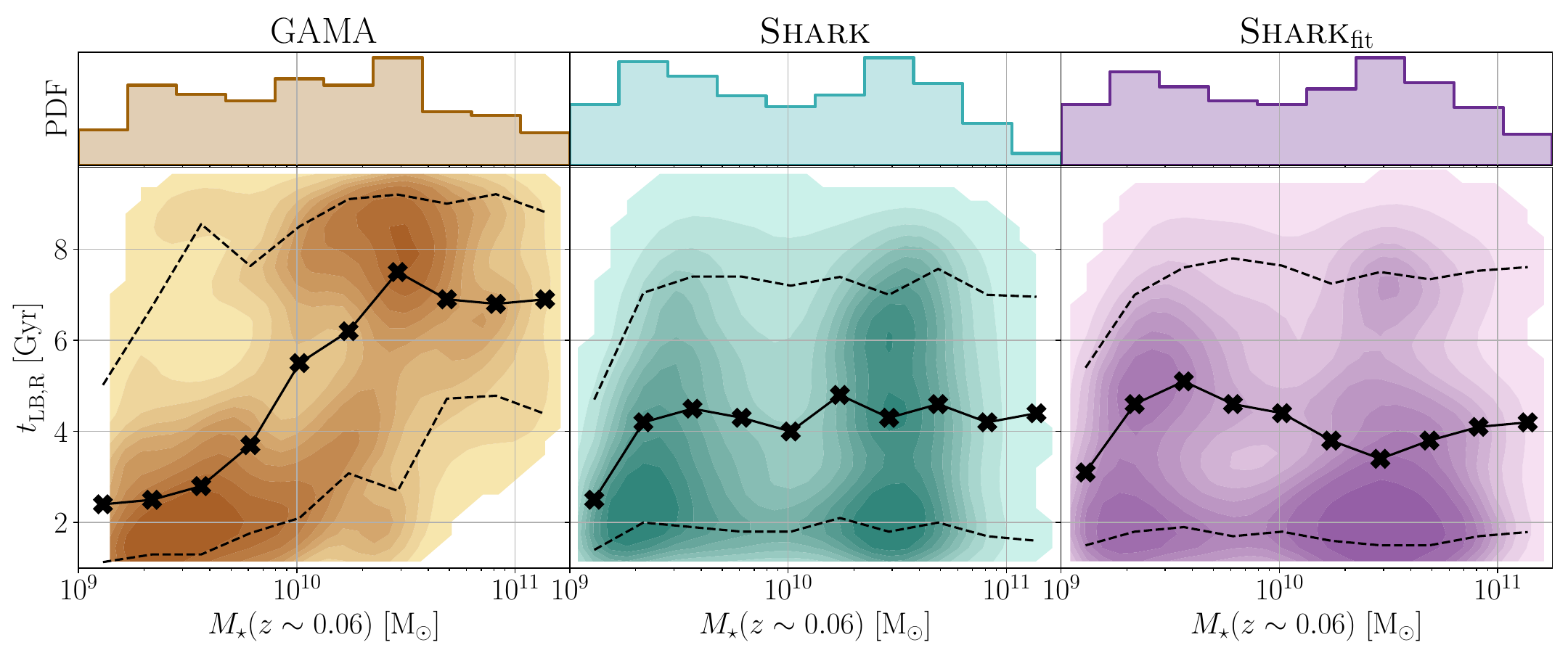}
    \caption{The lookback time when (currently red, $z\sim0.06$) galaxies joined the red population (\tLBR) as a function of current stellar mass, together with the stellar mass distribution, for each of our samples selected following Table \ref{tab:prob_dist}.
    For each individual sample, the solid lines indicates the \tLBR\ running median, the dashed lines the running 16-84$\mathrm{th}$ percentiles, both using the same bins as the stellar mass histograms of the top panels.
    The black markers indicate the stellar mass and \tLBR\ median point for each bin.
    The background contours indicate the smooth distribution obtained using the \texttt{gaussian\_kde} Gaussian Kernel Density Estimator (KDE) function from \textsc{scipy}, to avoid visualisation artefacts due to the discreteness of our data in \tLBR--\tauQ\ space.
    The lightest contour shows the highest-density region containing 99\% of the mass of the Gaussian KDE, with the rest of the contours evenly spaced in percentage of the mass contained.}
    \label{fig:all_Mstar_tLBR}
\end{figure*}

We explore the fractions of galaxies in blue/transitional/red regions across cosmic time in Figure \ref{fig:popfrac_censat}.
We also show in Figure \ref{fig:popfrac_censat} the fractions for both central and satellite galaxies, according to their classification at observation time.
For this, we follow the same convention adopted in \citet{bravo2020} of treating all isolated and central group galaxies\footnote{Those with $\texttt{RankIterCen}=1$ in the GAMA Galaxy Group Catalogue, from the iterative ranking procedure defined in section 4.2.1 of \citet{robotham2011a}.} from GAMA as centrals, and the remaining galaxies as satellites.
Since we demonstrated in that work that central/satellite confusion plays is an important factor, we show the results for both the true central/satellite classification in \shark/\sharkfit\ and a confused classification.
We note that we use a higher level of confusion than in \citet{bravo2020}, 23\% instead of 15\%, because the sample we use from GAMA in this work is limited to a significantly lower redshift ($z<0.06$ instead of $z<0.6$).
This elevated confusion can be seen in figure 3 of \citet{bravo2020}, and we first presented and tested this higher value in \citet{chauhan2021}.

The time dependence (independence) of the density for intermediate values of \PR\ seen for GAMA and \sharkfit\ (\shark) in Figure \ref{fig:PRdist_bymass&time} is clearly reflected in the time evolution of the transitional fraction of galaxies shown in Figure \ref{fig:popfrac_censat}.
The red fraction is in excellent agreement between \shark\ and \sharkfit, which indicates that we are accurately recovering this fraction with \prospect.
In contrast, the transitional fractions in \sharkfit\ are in better agreement with GAMA than \shark, suggesting that this is to some degree affected by the modelling choices in \prospect.
We note that the increased blue fraction in \sharkfit\ is consistent with the delayed SFHs relative to those of \shark\ that we showed in appendix A of \citetalias{bravo2022}, an issue we found related to a hard-to-solve degeneracy between dust and star formation parameters for bulge-dominated massive galaxies in \shark.

Central galaxies show a higher blue fraction than satellites in all samples, but there are differences across samples.
\shark\ and \sharkfit\ predict a higher fraction of blue centrals relative to GAMA, at lookback times of $\lesssim5$ Gyr for the former and at all lookback times for the latter, though including central/satellite classification confusion lessens this tension.
The opposite trend is true for the red population, being under-estimated by \shark/\sharkfit\ compared to GAMA.
Interestingly, while \sharkfit\ shows a strong under-prediction of the transitional fraction of centrals at $\gtrsim3$ Gyr relative to \shark, the difference is mostly absorbed by the blue fraction, which is overestimated (underestimated) in \sharkfit\ above (below) a lookback time of $\sim2$ Gyr.
This points to the \tLBB\ recovered for centrals being biased towards later times.

\shark\ and \sharkfit\ exhibit a significantly higher fraction of red satellites compared to GAMA, reaching $\sim80\%$ at $t^{}_\mathrm{LB}=1$ Gyr, a factor of $\sim3$ larger than observations, but this tension is strongly reduced when accounting for central/satellite classification confusion.
The transitional satellites in \sharkfit\ show a similar difference to those in \shark\ as previously mentioned for centrals, but unlike centrals, the under-abundance of transitional satellites in \sharkfit\ is balanced out by an over-abundance of both blue and red satellites.
This difference suggests that the SFH model parameterisation adopted in \prospect\ may cause the transition measured to be too fast.
GAMA and \sharkfit\ exhibit a qualitatively similar evolution for the transition fraction, and they come into quantitative agreement at lookback times of $\gtrsim6$ Gyr, in line with the results from \citetalias{bravo2022}.

\section{Distribution and time evolution of \tauQ}\label{sec:timescales}

\begin{figure*}
    \centering
    \includegraphics[width=\linewidth]{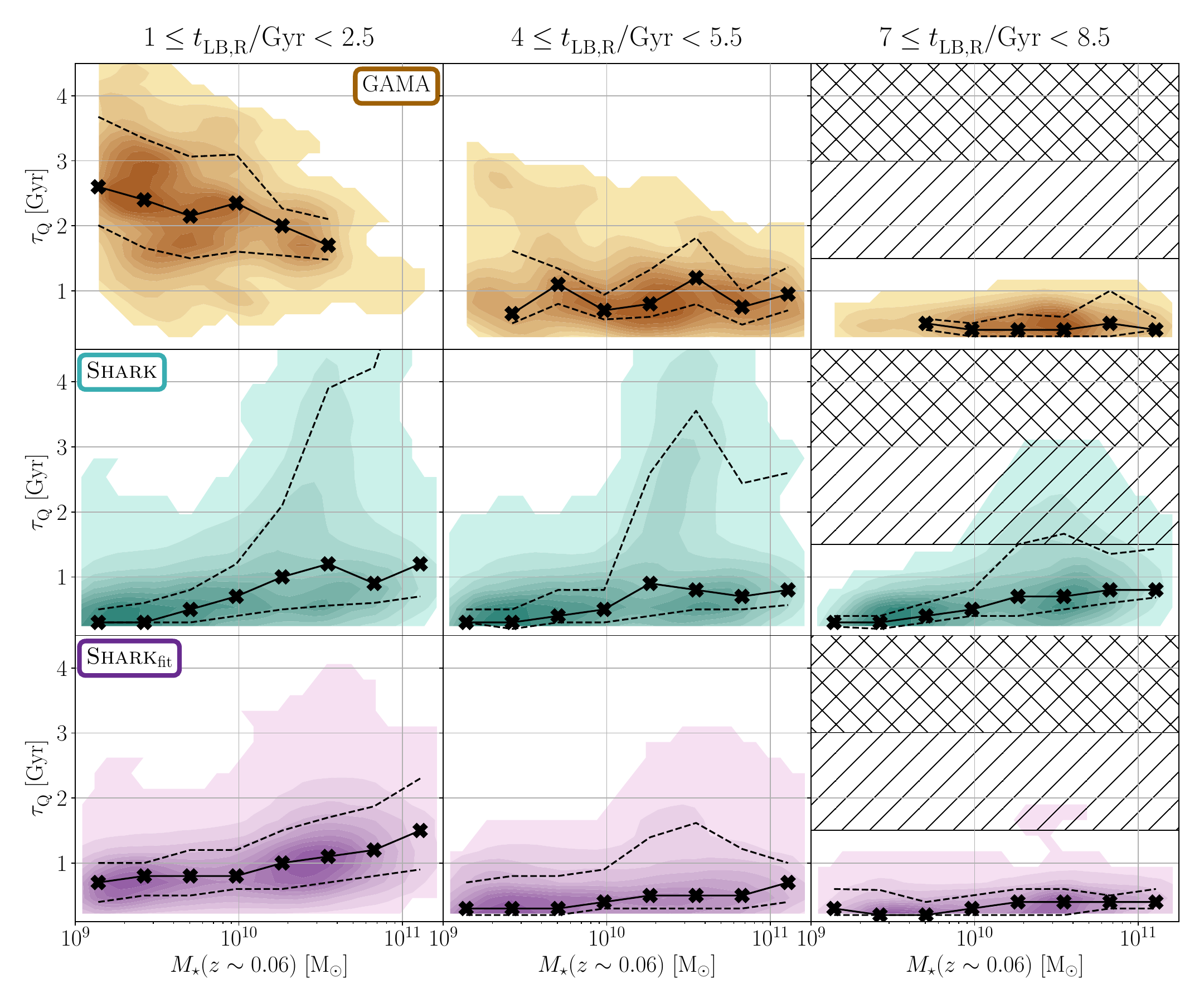}
    \caption{Blue-to-red transition timescales (\tauQ) as a function of current stellar mass.
    The distributions are shown for three \tLBR\ bins: $1\leq$\tLBR$<2.5$ (left column), $4\leq$\tLBR$<5.5$ (middle column), and $7\leq$\tLBR$<8.5$ (right column).
    Each row corresponds to a different sample, from top to bottom: GAMA, \shark, and \sharkfit.
    Solid lines, dashed lines, markers and contours as in Figure \ref{fig:all_Mstar_tLBR}.
    The diagonally-hatched region indicates where the \tauQ\ measurements become incomplete in the corresponding \tLBR\ bin, and the cross-hatched where no \tauQ\ measurement is possible (only visible on the right-most column due to our choice of limits for the $y$-axis).
    Note that the \tLBR\ bin shown in the left column lies in the range of lookback times that we found affected by SED-fitting-related biases in \citetalias{bravo2022}.}
    \label{fig:all_Mstar_tauB2R}
\end{figure*}

Defining both \tLBB\ and \tLBR\ is straightforward for GAMA and \sharkfit, as \PR\ is monotonically-increasing due to our choice of SFH in \prospect, with \tLBB\ (\tLBR) simply being the last (first) time the galaxy was a member of the blue (red) population.
The caveat to this statement is that, depending on the details of the evolution of the galaxy colour populations as a whole, a galaxy may see a decrease in \PR\ without changing its colour.
We do observe this behaviour in both GAMA and \sharkfit, being particularly clear for galaxies above $\sim$\mstar{10.5} in the former.
For this reason, we force a monotonic time evolution of \PR\ for galaxies that cross our \PR$=0.98$ threshold in these two samples.
We find the maximum \PR\ for all galaxies after they become red, and then set all subsequent values of \PR\ to this maximum value.

\begin{figure*}
    \centering
    \includegraphics[width=\linewidth]{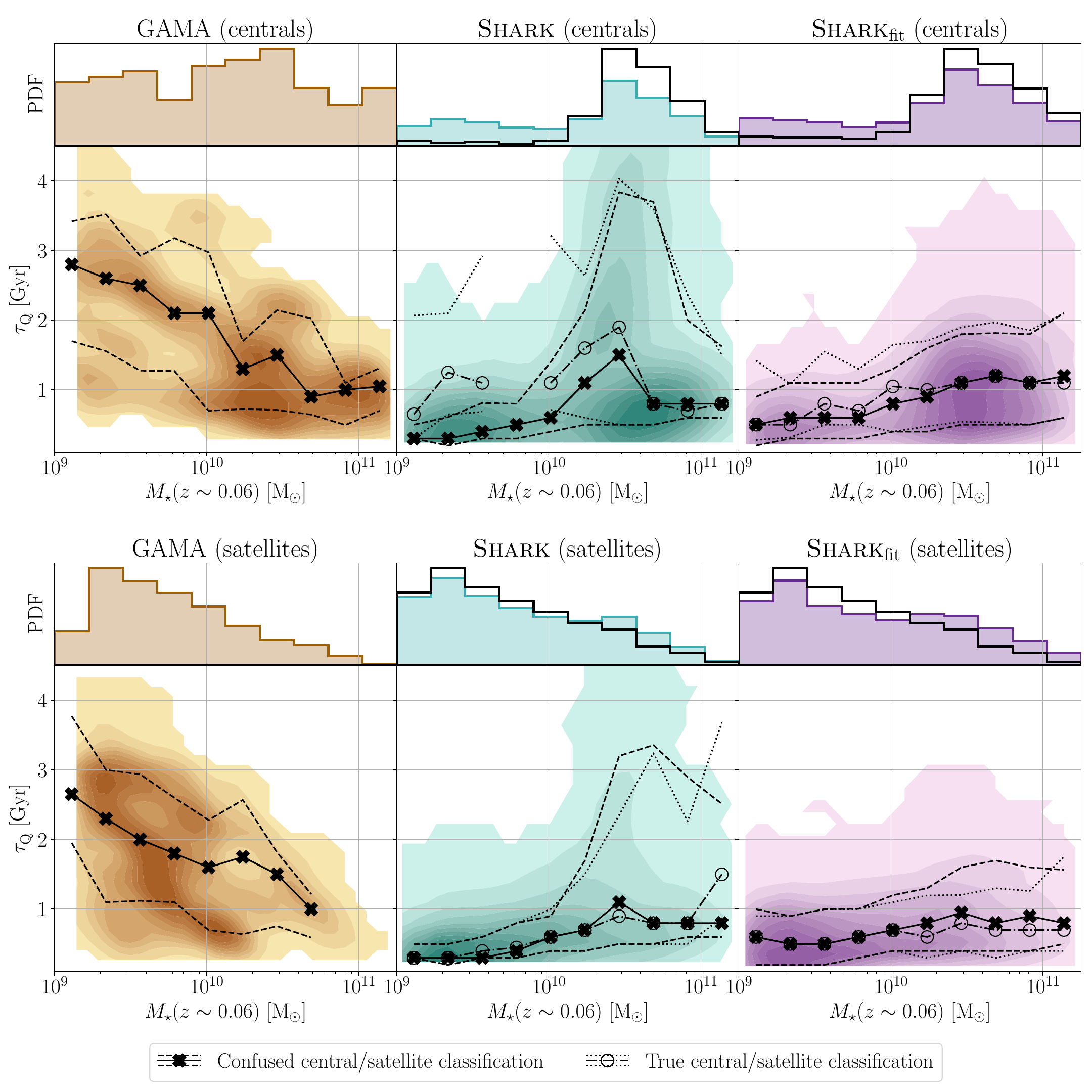}
    \caption{\tauQ\ as a function of current stellar mass, together with the stellar mass distribution, divided between centrals and satellites.
    Only galaxies with $1\leq$\tLBR$<6$ Gyr are shown.
    For each individual sample, the solid/dashed (dash-dotted/dotted) lines indicate the \tauQ\ running median/16-84$\mathrm{th}$ percentiles for the GAMA-like (true) central-satellite classification, both using the same bins as the stellar mass histograms.
    The black cross (open) markers indicate the stellar mass and \tLBR\ median point for the GAMA-like (true) central-satellite classification in each bin.
    Contours as in Figure \ref{fig:all_Mstar_tLBR}.
    The coloured (black) histograms show the measured stellar mass distribution for the GAMA-like (true) central-satellite classification.
    Note that the gap seen in the running median for the true centrals in \shark\ corresponds to a mass bin where no galaxies are present (see histogram above).}
    \label{fig:censat_Mstar_tauB2R}
\end{figure*}

\shark\ SFHs are not forced to be monotonic, and while rejuvenation in \shark\ is not a common occurrence in general ($\lesssim3$\%, see Appendix A2 of \citetalias{bravo2022}), the measurement of \tauQ\ for galaxies that do rejuvenate presents a challenge (mostly massive centrals).
To measure \tauQ\ in \shark, we first select galaxies that are blue at $t^{}_\mathrm{LB}=10$ Gyr and that are red $t^{}_\mathrm{LB}=1$ Gyr, trace the continuous time span during which the galaxy was red, and the latest time before this period that the galaxy was blue (see Appendix \ref{app:recovery} for a discussion on different definitions).
Table \ref{tab:prob_dist} shows the fraction of red galaxies in our three samples, indicating that selecting galaxies being blue at least at a lookback time of 10 Gyr discards $\sim$20--40\% of the current red galaxies, with GAMA seeing the largest reduction in sample size and \shark\ the smallest.

Before we explore the measured \tauQ\ from our samples, we first discuss the \tLBR\ distributions in Figure \ref{fig:all_Mstar_tLBR}.
While all three samples display roughly similar stellar mass distributions for red galaxies (though \shark/\sharkfit\ show a bimodality not clear in GAMA), there are differences in when these galaxies became red.
GAMA exhibits a clear trend in \tLBR\ with stellar mass, with more massive galaxies becoming red at earlier times.
In contrast, the \tLBR\ distribution in both \shark\ and \sharkfit\ are broadly consistent for stellar masses above $\gtrsim9.3$ Gyr , i.e., the time when \shark\ galaxies become red shows no clear trend with stellar mass.
The overall good agreement between both \shark\ and \sharkfit\ indicates that there are no strong biases in our GAMA measurements, hence the strong difference between GAMA and \shark/\sharkfit\ is not a consequence of our colour evolution reconstruction.
In other words, the fact that we do not find a downsizing trend in \shark/\sharkfit\ as strong as in GAMA is a short-coming of the physical models in \shark\footnote{The reader may find this tension to resemble the historical one of too many star-forming massive galaxies addressed in the works of, e.g., \citet{bower2006,cattaneo2006,croton2006}.
    These works proposed solutions to (then current) tension in the fraction of $z\sim0$ massive passive galaxies, i.e., \textit{how many} massive red galaxies the early galaxy formation models.
    The tension we find in this work is in the \tLBB\ distribution of $z\sim0$ massive passive galaxies, i.e., \textit{when} massive passive galaxies became red.
    Hence, our findings showcase the need for second-order corrections to the modelling of galaxy formation relative to those presented in \citet{bower2006,cattaneo2006,croton2006}.}

\subsection{\tauQ\ distribution of the overall galaxy population}\label{subsec:all}

In Figure \ref{fig:all_Mstar_tauB2R} we show the distribution of \tauQ\ as a function of stellar mass, divided into three lookback time bins.
The distributions are roughly consistent across cosmic time in \shark, the largest difference being the increased dispersion above $\sim$\mstar{10}, which suggests that there is no strong time evolution of the \tauQ\ distribution.
In contrast, the \tauQ\ distribution of GAMA changes as a function of \tLBR, with the median \tauQ\ increasing by a factor of $\sim3$ in the span of 4.5 Gyr.

GAMA and \sharkfit\ display similar timescale-mass relations in the highest \tLBR\ bin.
This is in line with the results in \citetalias{bravo2022}, where we found a strong similarity in the galaxy distributions of both GAMA and \sharkfit\ in colour-mass space at high lookback times ($t^{}_\mathrm{LB}>6$ Gyr), likely driven by dust parameter degeneracies (see Appendix A of \citetalias{bravo2022} for further details).
The timescale-mass relations measured in \shark\ and \sharkfit\ are in good agreement in the other two lookback time bins, indicating that we can recover this with \prospect, which validates the difference between both and GAMA as real.
The only significant difference between \shark\ and \sharkfit\ at these lower lookback times is that we do not recover the longest \tauQ\ from \shark, possibly due to episodes of weak rejuvenation extending the time period galaxies remain in the transitional region.
The \tauQ--$M_\star$ relation that we observe in GAMA is in clear tension with that we predict in \shark, which also exhibit opposite trends with stellar mass, i.e., low-mass galaxies in GAMA take much longer to quench at recent times than in \shark.
We explore the effect of selection biases in the measured \tauQ\ evolution in Appendix \ref{app:timescale_evol}, where we find that GAMA does exhibit a strong time evolution of the \tauQ\ distribution, a mild evolution in \sharkfit, and that the \tauQ\ distribution in \shark\ is independent of cosmic time.

\begin{figure*}
    \centering
    \includegraphics[width=0.98\linewidth]{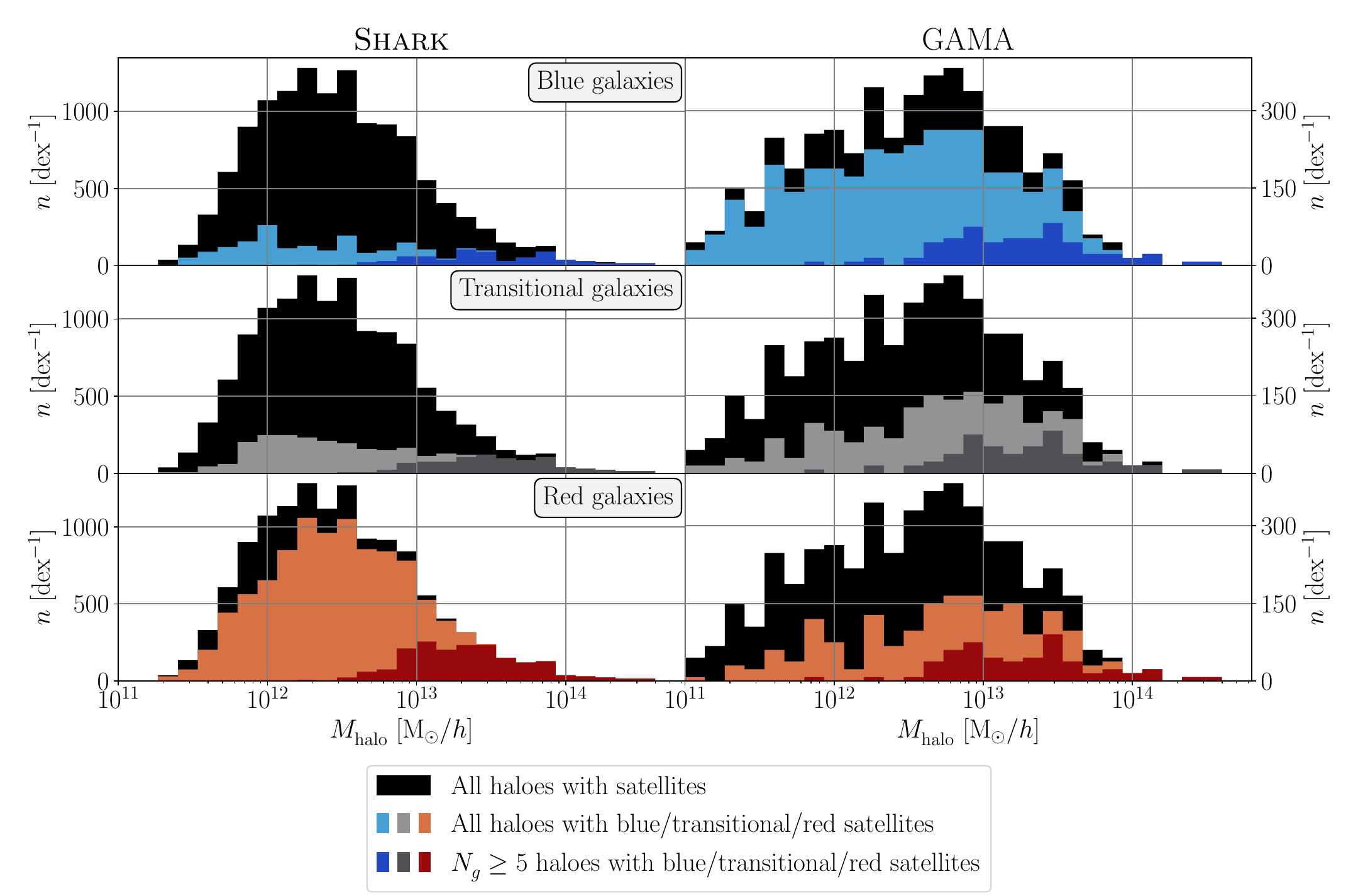}
    \caption{Comparison of the halo mass distribution from the GAMA and \shark\ samples.
    \shark\ is shown on the left column and GAMA on the right.
    In all panels, the black histograms indicate the distribution of all haloes hosting at least one satellite galaxy.
    The coloured histograms in the top/middle/bottom row show the mass distribution of haloes hosting at least one blue/transitional/red satellite.
    The lighter-coloured histograms show the distribution for all said haloes hosting at least one satellite, with the darker-coloured histograms the subset that hosts at least four satellites ($N_g\geq5$).
    Note that both panels have different scales for the y-axis to focus on the effect of each selection rather the different sizes of our samples.}
    \label{fig:halodist}
\end{figure*}

\subsection{Environmental effects on \tauQ}\label{subsec:env}

We now explore how \tauQ\ compares between central and satellite galaxies.
Figure \ref{fig:censat_Mstar_tauB2R} shows both the stellar mass distribution and \tauQ\ as a function of their current stellar mass, divided into centrals and satellites.
For \shark\ and \sharkfit\ we show the results for both true and GAMA-like central/satellite classifications.

True centrals and satellites in \shark\ show a significant difference in \tauQ\ for galaxies below $\sim$\mstar{10.5}, with centrals exhibiting longer timescales than satellites.
Satellites in \sharkfit\ are in good agreement with those from \shark.
Centrals show less of an agreement for \tauQ, in particular the timescales in \sharkfit\ for centrals of $\sim$\mstar{9.5} ($\sim$\mstar{10.5}) are shorter than in \shark\ by a factor of $\sim4$ ($\sim2$).
In comparison, GAMA centrals and satellites only differ at $M_\star\lesssim$\mstar{10}, with \tauQ\ of satellites being $\sim0.4$ Gyr than for centrals.
The difference between centrals and satellites is reduced when using a GAMA-like classification for both \shark\ and \sharkfit, suggesting the possibility of a larger difference between GAMA centrals and satellites than what we measure.

To further explore the differences for satellites between GAMA and \shark, Figure \ref{fig:halodist} shows the mass distribution of haloes that host satellites in both samples.
To quantify the differences between GAMA and \shark, Table \ref{tab:WD} contains the first Wasserstein distances between the several selections shown in Figure \ref{fig:halodist}.
\citet{chauhan2021} found that the recovery of \shark\ halo masses with the \citet{robotham2011a} group finder, which was used to infer halo masses for GAMA, is reasonable for the high-multiplicity groups ($N_g\geq5$) but the quality of the recovery noticeably decreases for low-multiplicity groups.
Those results can account for the difference between GAMA and \shark\ in the distribution of halo masses for haloes hosting at least one satellite.
The GAMA and \shark\ halo mass distributions for all groups hosting at least one satellite are significantly different, irrespective of the \PR\ selection applied to both samples, with similar first Wasserstein distances.
The relative distribution of blue/red galaxies are also different, with the majority of haloes in \shark\ that host at least one satellite also hosting at least one red satellite, in strong contrast to GAMA.

Figure \ref{fig:halodist} and Table \ref{tab:WD} show that the halo mass distribution of $N_g\geq5$ groups with red satellites in GAMA and \shark\ are significantly closer, hence this should be a strong probe for the treatment and evolution of the satellites residing in such haloes.
For centrals, we ignore $N_g\geq5$ group red centrals, as there are $\lesssim$40 of the latter in GAMA and \shark\footnote{There are significantly more in \sharkfit, factor of $\sim4$ more than in \shark, but this is a consequence of the delayed formation of the red population from our \prospect\ fits to \shark.
    This could be a result of the relative fractions of galaxies that undergo rejuvenation, as defined in \citetalias{bravo2022}, as $\sim20\%$ of the red centrals in \shark\ had at least one rejuvenation episode, compared to only $\sim4\%$ of satellites.
    We should also note that we are using a subset of the full simulation box for \shark/\sharkfit, so it is possible to increase this number by a factor of $\sim$30.
    The low number from GAMA is the limit to study $N_g\geq5$ group red centrals.}
.
Overall, we find little difference between true central (all true satellite) and isolated ($N_g\geq5$ satellite) galaxies in all three samples, with the most important finding being that \shark/\sharkfit\ lack the observed numbers of $\lesssim$\mstar{10} red isolated galaxies we find in GAMA (see Figure included in the supplemental material).

\begin{table}
    \centering
    \begin{tabular}{l|c}
        Selection & Wasserstein distance \\
        \hline
        All haloes & 0.275 \\
        All haloes, blue satellites & 0.306 \\
        All haloes, transitional satellites & 0.231 \\
        All haloes, red satellites & 0.267 \\
        $N_g\geq5$ haloes, blue satellites & 0.246 \\
        $N_g\geq5$ haloes, transitional satellites & 0.267 \\
        $N_g\geq5$ haloes, red satellites & 0.105 \\
    \end{tabular}
    \caption{The Wasserstein distance between the each corresponding pair for the GAMA and \shark\ halo mass distributions shown in Figure \ref{fig:halodist}.}
    \label{tab:WD}
\end{table}

\begin{figure*}
    \centering
    \includegraphics[width=\linewidth]{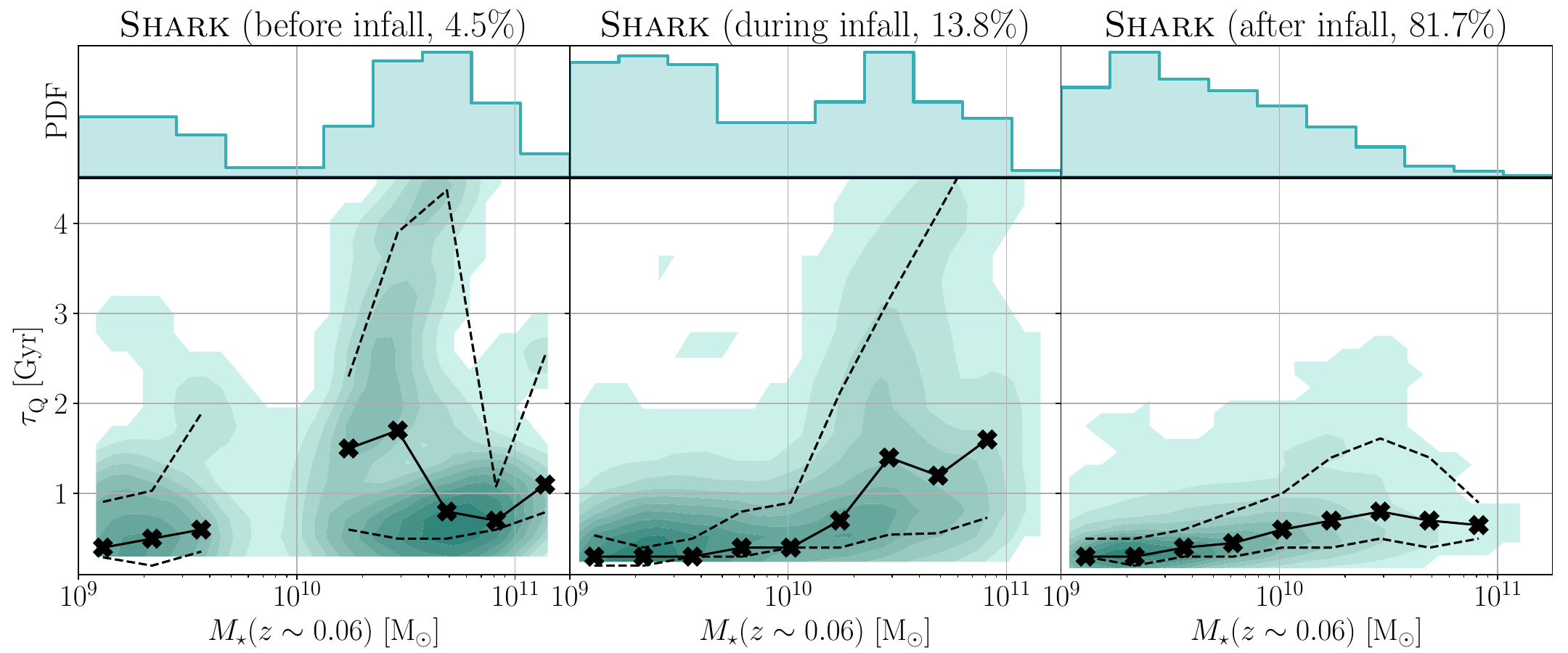}
    \caption{\tauQ\ of satellites in \shark\ as a function of current stellar mass, and the stellar mass distribution, divided by when a galaxy became red relative to the time when the galaxy became a satellite.
    The percentage that each sample represents of the total of red satellites in \shark\ is shown on the top labels of each column.
    Solid lines, dashed lines, markers and contours as in Figure \ref{fig:all_Mstar_tLBR}.}
    \label{fig:infall_Mstar_tB2R}
\end{figure*}

\subsection{Connection between \tauQ\ and satellite infall in \shark}\label{subsec:shark_infall}

One of the powerful features of using simulations is that they enable us to further explore the evolution of red galaxies.
In particular, here we explore how \tauQ\ is linked to the central-to-satellite transition in \shark, with which we explore the effect of implemented physical models in driving \tauQ.
While we can also explore this aspect with \sharkfit, Appendix \ref{app:recovery} shows that there are reasons not to.
Both \tLBB\ and \tLBR\ are not as well-recovered as \tauQ, which leads to significant confusion on whether a satellite became red before, during, or after infall, e.g., the percentage of the former increases from $\sim4\%$ to $\sim15\%$.
Classifying whether a \sharkfit\ satellite became red based on \tLBR\ measured in \shark\ would limit our analysis to how well we recover the evolution of centrals and satellites (see Appendix \ref{app:recovery}).
For GAMA, this effect would be further compounded by the relatively limited ability to extract infall times from observations\footnote{E.g., the method outlined in \citet{pasquali2019} divides phase-space into a series of distinct regions characterised by a monotonically-increasing mean time of infall.
    The main limitation of such methods is the significant overlap in infall times ranges between zones.
    Furthermore, we note to the reader that \tauQ, $\Delta$\tLBB\ and $\Delta$\tLBR, and the scatter in the infall time found by \citet{pasquali2019} are all in the order of $\sim2$ Gyr, which we expect to strongly obscure the correlation between infall and quenching, if present.}

We classify satellite galaxies in \shark\ in three groups, based on the lookback time of infall ($t^{}_\mathrm{LB,infall}$) relative to its transition from blue to red, those that:
\begin{enumerate}
    \item became red before they became a satellite, i.e., transitioned in colour before infall, \tLBB$>$\tLBR$>t^{}_\mathrm{LB,infall}$;\label{item:SharkBI}
    \item were a central the last time they were blue, but were a satellite by the time they became red, i.e., infall happened during the colour transition, \tLBB$>t^{}_\mathrm{LB,infall}>$\tLBR;\label{item:SharkDI}
    \item were still blue when they became a satellite, i.e., transitioned in colour after infall, $t^{}_\mathrm{LB,infall}>$\tLBB$>$\tLBR.\label{item:SharkAI}
\end{enumerate}

The distribution of stellar mass and \tauQ\ for these categories are shown in Figure \ref{fig:infall_Mstar_tB2R}.
It is clear that becoming a satellite is the main driver for galaxies to become red, as 81.7\% of the red satellites in \shark\ became red before after infall \ref{item:SharkAI}.
The stellar mass distribution of galaxies that became red before \ref{item:SharkBI} and after infall show a marked difference, with those that became a satellite while in transition \ref{item:SharkDI} showing a distribution intermediate between the other two.
\shark\ galaxies galaxies that became red after infall have the shortest \tauQ, while those galaxies that became red before infall have the longest \tauQ.
Galaxies that became a satellite while in transition show intermediate \tauQ\ relative to the other two groups.

\section{Discussion}\label{sec:disc}

\subsection{Comparing our \tauQ\ definition with previous literature}\label{subsec:our_selections}

\begin{table*}
    \centering
    \begin{tabular}{c|c|c|c|c}
        Reference & Data type & Parameter space & Transition region lower limit & Transition region upper limit \\
        \hline
        \citet{schawinski2014} & Observation & $(u-r)$--$M_\star$ & $0.25\log_{10}(M_\star/\mathrm{M}_\odot)-0.75$ & $0.25\log_{10}(M_\star/\mathrm{M}_\odot)-0.24$ \\
        \citet{smethurst2015} & Observation & $(u-r)$--$r$ & $-0.244\mathrm{tanh}\left(\frac{r+20.07}{1.09}\right)+20.6-\sigma$ & $-0.244\mathrm{tanh}\left(\frac{r+20.07}{1.09}\right)+20.6+\sigma$ \\
        \citet{trayford2016} & Simulation & $(u-r)$--$M_\star$ & $0.2\log_{10}(M_\star/\mathrm{M}_\odot)-0.25z^{0.6}-0.3$ & $0.2\log_{10}(M_\star/\mathrm{M}_\odot)-0.25z^{0.6}+0.24$ \\
        \citet{bremer2018} & Observation & $(u-r)$--$M_\star$ & $0.1\log_{10}(M_\star/\mathrm{M}_\odot)+0.3$ & $0.2\log_{10}(M_\star/\mathrm{M}_\odot)-0.5$ \\
        \citet{nelson2018a} & Simulation & $(g-r)$--$M_\star$ & $\mu^{}_\mathrm{B}+\sigma^{}_\mathrm{B}$ & $\mu^{}_\mathrm{R}-\sigma^{}_\mathrm{R}$ \\
        \citet{phillipps2019} & Observation & $(u-r)$--$M_\star$ & $0.1\log_{10}(M_\star/\mathrm{M}_\odot)+0.3$ & $0.2\log_{10}(M_\star/\mathrm{M}_\odot)-0.5$ \\
        \citet{wright2019} & Simulation & $(u-r)$--$M_\star$ & $\mu^{}_\mathrm{B}+1.5\sigma^{}_\mathrm{B}$ & $\mu^{}_\mathrm{R}-1.5\sigma^{}_\mathrm{R}$ \\
        This work & Both & $(u-r)$--$M_\star$ & \PR$>0.02$ & \PR$<0.98$
    \end{tabular}
    \caption{Sample of literature criteria to define the green valley/transition region.
    Our classification is at the end of the table for comparison purposes.
    Several remarks need to be made for a fair comparison.
    All literature definitions using observations include no time evolution and are valid only at low redshift ($z\leq0.25$).
    The definition by \citet{smethurst2015} references a dispersion ($\sigma$), but it is not clear what dispersion they used, besides that it seems to be independent of stellar mass (see their figure 3).
    \citet{bremer2018} and \citet{phillipps2019} use the same definition, but the former limits it to a narrow stellar mass range ($10^{10.25}<M\star/\mathrm{M}_\odot<10^{10.75}$), while the latter expands it to their full sample.
    \citet{nelson2018a} and \citet{wright2019} use the same type of criteria, but they chose difference factors for the standard deviation, whether this is a consequence of their different choices for colour or not is not clear.
    Note that $\mu^{}_\mathrm{\{B,R\}}$, $\sigma^{}_\mathrm{\{B,R\}}$, and \PR\ all are functions of lookback time, stellar mass and colour, see Section \ref{sec:probabilities} for details.}
    \label{tab:lit}
\end{table*}

\begin{figure*}
    \centering
    \includegraphics[width=\linewidth]{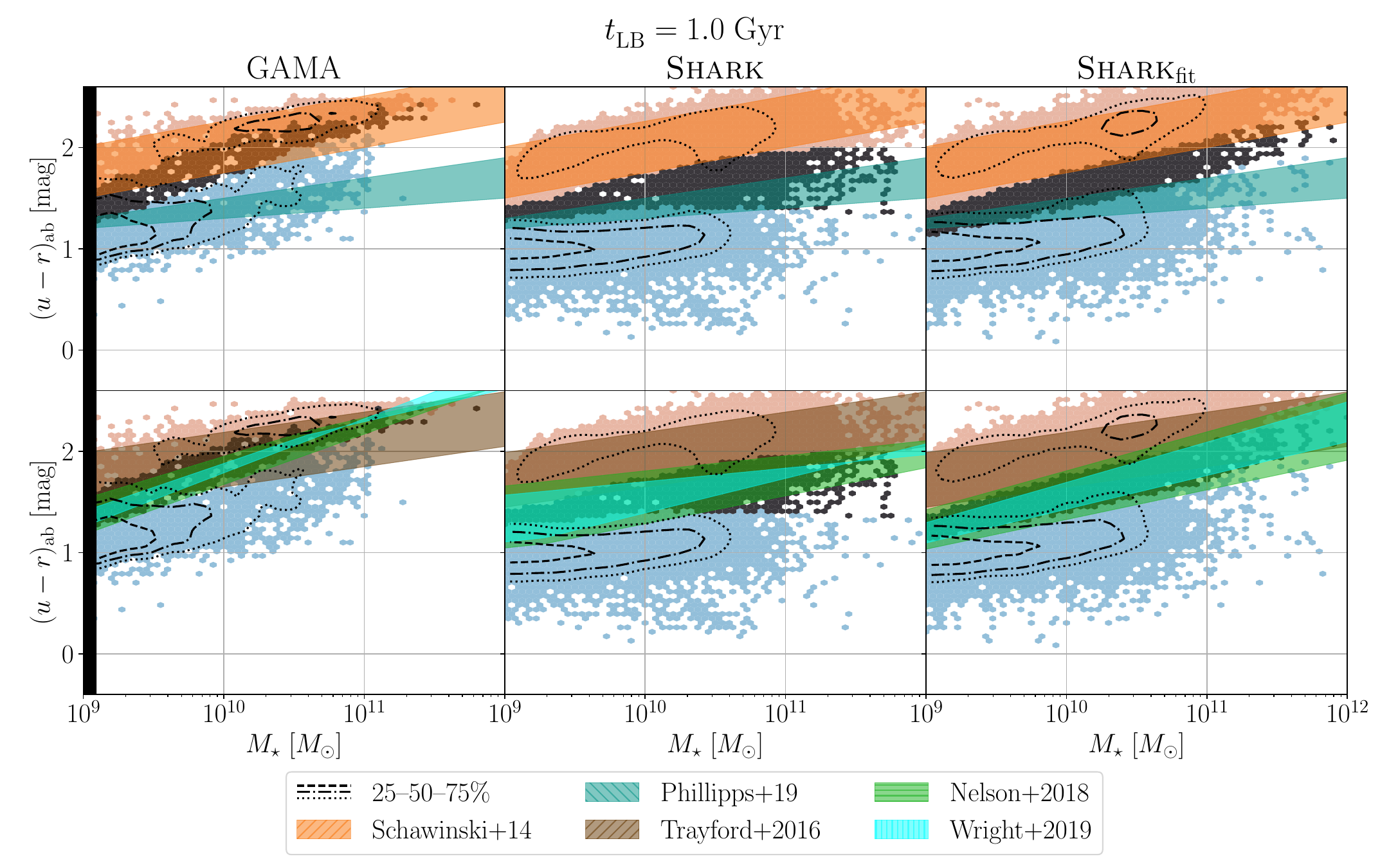}
    \caption{Comparison of the colour classification we adopt to two observational \citep[, top row; in orange and teal, respectively]{schawinski2014,phillipps2019} and three theoretical \citep[, bottom row; in brown, green, and cyan, respectively]{trayford2016,nelson2018a,wright2019} literature examples.
    The comparison is made at the lowest lookback time of our data (1 Gyr), as a rough middle point between the redshift ranges of the literature classifications.
    Each column shows the classification for one of our samples, following the same thresholds as in Figure \ref{fig:PRdist}, with the underlying histogram bins being coloured in blue/grey/red if they median \PR\ of the galaxies classifies them as being blue/transitional/red.
    The dashed/dash-dotted/dotted contours indicate the highest density regions in each panel that contain 25/50/75\% of the galaxies of each sample.
    The transitional region from the literature examples are shown by coloured bands in each panel.
    The black region in the left column indicates the stellar mass below the stellar mass completeness limit for GAMA (lies below \mstar{9} for \shark\ and \sharkfit, further details in section 3.2 of \citetalias{bravo2022}).
    The classifications from \citet{nelson2018a} and \citet[]{wright2019} are defined as functions of the means and standard deviations of each population, which is the reason why they are different across our samples.}
    \label{fig:PRlit_tLBR_Mstar}
\end{figure*}

Fundamental to any measurement of \tauQ\ is how it is defined.
Measurements of \tauQ\ in the literature include derivation from star formation rates \citep[e.g.,][]{wetzel2013,belli2019,tacchella2022}, galaxy colours \citep[e.g.,][]{schawinski2014,trayford2016,bremer2018,mcnab2021}, and spectral properties \citep[e.g.,][]{wheeler2014,rowlands2018,angthopo2019}.
Definitions include timescales to cross specific thresholds \citep[e.g.,][]{trayford2016,bremer2018,tacchella2022}, $e$-folding timescales \citep[e.g.,][]{wetzel2013,schawinski2014,wheeler2014,bremer2018}, and inference from population densities or ages \citep[e.g.,][]{rowlands2018,angthopo2020,mcnab2021}.
Establishing how these different measurements compare is outside the scope of this work, so we only compare our \tauQ\ definition and measurements to those in the literature that are derived from galaxy colours.
We remind the reader that differences in \tauQ\ definitions are not the only challenge in performing these comparisons
Modelling differences and shortcomings in the recovery of the true intrinsic galaxy colours can easily permeate both \tauQ\ definitions and measurements, so comparisons with literature results are also limited by underlying modelling differences.\footnote{It is worth noting that approaches based on relatively dust-insensitive spectral indexes \citep[e.g.,][]{rowlands2018,angthopo2019} provide a different method to measure \tauQ.
    We do not include such studies in our comparisons as they are significantly different in at least one key aspect, like chosen space compared to our choice of $u-r$--$M_\star$ \citep[e.g.,][]{rowlands2018} or in the definition of the transitional region \citep[e.g.,][]{angthopo2019}.}

While definitions from observations abound \citep[e.g.,][]{schawinski2014,smethurst2015,bremer2018,phillipps2019}, differences in the recovery of the intrinsic stellar light result in significant differences in the loci of the colour populations.
Table \ref{tab:lit} provides a description of these selections, and the top row of Figure \ref{fig:PRlit_tLBR_Mstar} shows how those from \citet{schawinski2014} and \citet{bremer2018,phillipps2019} compare to ours.
The issue of definitions that adopt a fixed parameterisation instead of being a function of population properties are clear here as the match to ours is strongly sample-driven.
The \citet{schawinski2014} green valley selection covers almost exclusively the red population for \shark/\sharkfit, and the \citet{phillipps2019} covering mostly the blue population for GAMA\footnote{This is not to say that these are poor selections for the samples for which they were designed \citep[see figures 4 and 1 of][, respectively]{schawinski2014,phillipps2019}, which we could only assess by implementing our method to their data sets.}.
Furthermore, (to the author's knowledge) there are no observational selections that account for colour evolution with cosmic time, i.e., the selections are at fixed lookback time/redshift.

We now compare the classifications used for simulations by \citet{trayford2016,nelson2018a,wright2019} to the one we adopt for this work, shown in the bottom row of Figure \ref{fig:PRlit_tLBR_Mstar}.
\citet{trayford2016} classifies galaxies from the EAGLE simulation \citep{schaye2015} between red, green and blue using straight lines in the colour-mass plane, where only the normalisation evolves with time.
While it does overlap most of the region we classify as transitional in GAMA, it is clearly slanted as a function of stellar mass compared to our statistically-based selection.
The fixed nature of the selection limits also makes it a poor choice for \shark/\sharkfit, as it strongly overlaps the red population.
The \citet{trayford2016} selection limits are also consistently wider than our transitional region, save for \shark\ above $\sim$\mstar{10.5}, which would lead to an over-estimation of \tauQ.

The classifications from \citet{nelson2018a} for IllustrisTNG and \citet{wright2019} for EAGLE are more directly comparable to our classification, as they are all based on GMM fits to the colour population.
Both define their green selection as:
\begin{align}
    G_\mathrm{upper}&=\mu^{}_\mathrm{R}-f\sigma^{}_\mathrm{R},\\
    G_\mathrm{lower}&=\mu^{}_\mathrm{B}+f\sigma^{}_\mathrm{B},
\end{align}
\noindent where $\mu^{}_\mathrm{\{B,R\}}$ and $\sigma^{}_\mathrm{\{B,R\}}$ are the mean and standard deviation of the blue/red population, respectively, and $f$ is a constant value.
\citet{nelson2018a} adopts a value of $f=1$, while \citet{wright2019} adopts $f=1.5$.
For a fair comparison, we have used the $\mu^{}_\mathrm{\{B,R\}}$ and $\sigma^{}_\mathrm{\{B,R\}}$ we measured in \citetalias{bravo2022} to implement their selections, as the colour evolution in both EAGLE and IllustrisTNG may well not match those in GAMA and \shark.

Both are in reasonable agreement with our classification for \sharkfit, with \citet{nelson2018a} being a remarkable match, but the results for GAMA and \shark\ show that this is just a lucky coincidence.
In particular, in \shark\ both criteria fail to reproduce our probabilistically-based classification, and at low masses they lead to an over-estimation of the transitional region. 
At high masses both criteria under-estimate \tauQ.

\subsection{Comparing our \tauQ\ measurements between observations and simulations}\label{subsec:our_timescales}

\begin{figure*}
    \centering
    \includegraphics[width=\linewidth]{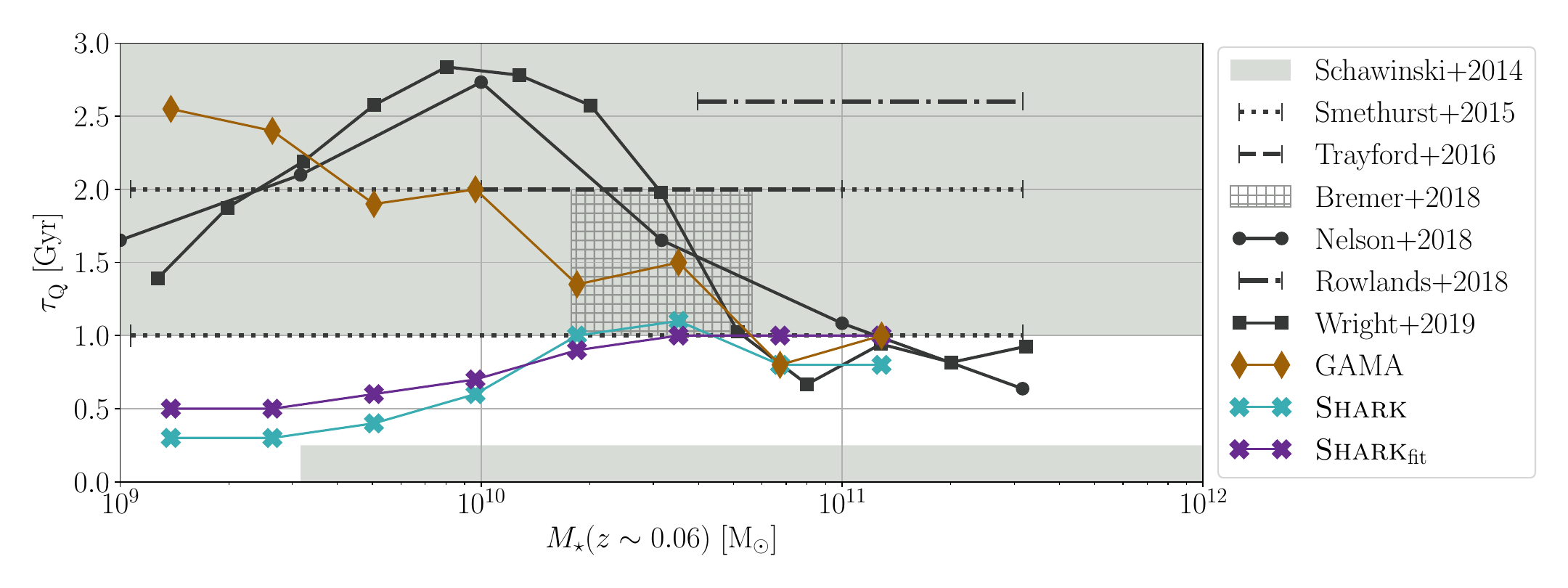}
    \caption{Comparison between the \tauQ\ we have measured in this work to a variety of literature results.
    Solid and hatched areas indicate results that only provide a range of values for \tauQ, together with the stellar mass range used in the work.
    Segmented lines indicate singular \tauQ\ values, with the extend of the line and markers on the edge the stellar mass range used to measure it.
    Solid lines indicate running medians of \tauQ\ as a function of stellar mass.
    Multiple lines may appear if the respective work provided more than one result.
    The results from this work are shown in colour, those from the literature in different shades of grey.
    Note that we show only one of the two measured \tauQ\ values from \citet{rowlands2018}, as the other lies outside our choice of range for \tauQ\ ($\sim6.6$ Gyr for $10^{11}<M_\star/$M$_\odot<10^{11.5}$)}
    \label{fig:final_fig}
\end{figure*}

A fundamental question to answer when exploring the \tauQ\ distribution is how it evolves with cosmic time.
This can provide an insight into the physical processes behind the colour transformation of galaxies.
A \tauQ\ distribution invariant with time would suggest that the different mechanisms and their relative prevalence are also time invariant.
If \tauQ\ evolves with cosmic time instead, that suggests some combination of the following: different mechanisms operate at different times, the mechanisms' efficiency change with time, or that the relative mix of these mechanisms evolves with time.

As described in Section \ref{sec:timescales}, the challenge in establishing whether \tauQ\ evolves with time is selection bias, irrespective of the chosen limits.
Furthermore, galaxies of different current stellar masses could have transitioned in colour at different lookback times.
When accounting for both lookback time and stellar mass dependencies, we find that GAMA shows a clear evolution towards longer timescales at more recent cosmic times, while the same is not evident in \shark.
We do find a time evolution for \tauQ\ in \sharkfit, unlike \shark, which suggests that the difference between GAMA and \shark\ may be attributable in to the process of SED fitting\footnote{More detailed exploration is required to discern whether this evolution is a product of the SED fitting process, and if so, whether this evolution is a reflection of the more limited imprint that older stellar populations have on galaxy SEDs or a limitation of one or more of the models implemented in \prospect.}.
The difference in evolution of \tauQ\ as a function of stellar mass between GAMA and \sharkfit\ suggests that some of this evolution might be real, instead of induced by our models in \prospect, though more work would be required to make a more conclusive statement.

Since the \tauQ\ distribution is broadly consistent for all three samples for \tLBR$<4$ Gyr, we can use that as a selection to study the connection between \tauQ\ and other galaxy properties.
This selection means that we are roughly halving the number of galaxies that we can explore in \shark/\sharkfit, as galaxies above $\sim$\mstar{9.3} exhibit a consistent median \tLBR\ of $\sim4$ Gyr in these samples, but for GAMA we will only be able to explore a small fraction of $\sim$\mstar{10} galaxies.
Comparing the \tauQ--$M_\star$ distribution for all three samples, shown in Figure \ref{fig:final_fig}, shows that \shark\ under-predicts \tauQ\ by up to $\sim2$ Gyr for $10^9$--\mstar{10.5}, with the tension being larger for lower stellar masses.
Since the \tauQ\ distribution in \sharkfit\ is in good agreement with that from \shark, and that for masses above $\sim$\mstar{10.5} all three samples show a good agreement, the difference between GAMA and \shark\ at lower masses suggests that the mechanisms that quench these lower mass galaxies in \shark\ are too efficient.

The red population in \shark\ is strongly dominated by satellites ($71.4\%$ of the sample), in apparent disagreement with GAMA ($50.8\%$), but as shown in \citetalias{bravo2022} this is alleviated by accounting for observational confusion between centrals and satellites.
Centrals and satellites below $\sim$\mstar{10.5} in \shark\ exhibit different \tauQ, but using a GAMA-like classification reduces the difference between both, suggesting that the similarity between centrals and satellites in GAMA could be due classification confusion.
The choice of central/satellite classification cannot account for the strong difference in \tauQ\ between GAMA and \shark/\sharkfit, as satellites exhibit shorter transition timescales in \shark/\sharkfit\ than GAMA.

Since $83.2\%$ of \shark\ satellites became red after infall, this points to the quenching mechanisms for satellites in \shark, the instantaneous stripping all halo gas of the galaxy, being too aggressive.
A more gradual stripping of gas would allow satellites to replenish their interstellar medium (ISM) gas, increasing \tauQ\ \citep[e.g.,][]{font2008}.
Other SAMs have adopted gradual halo gas stripping models, usually combined with the inclusion of ISM stripping to balance the availability of gas for satellites \citep[e.g.,][]{font2008,croton2006,croton2016,stevens2017,cora2018}.
More extreme models can also be found in the literature.
E.g., in \citet{henriques2015} satellites fully retain their halo gas and their ISM in haloes of masses lower than \mstar{14}, leading to slow exhaustion of the gas to star formation.
The latter was required by that model to reproduce the observed red fraction of galaxies.
Based on these results, we will explore in future work if the combination of gradual halo gas stripping and ISM stripping leads to longer timescales than instantaneous halo gas stripping without ISM stripping.

\shark\ satellites that became red before infall (i.e., that became red while being centrals) exhibit a similar \tauQ\ distribution as those of (current) central galaxies, but they show markedly different stellar mass distribution, with the former having a significant contribution from $\lesssim$\mstar{9.5} galaxies.
These low-mass satellites that became red as centrals exhibit starburst episodes prior to infall, indicating that the otherwise temporary gas exhaustion was made permanent by the instantaneous hot gas stripping in \shark.
Both low-mass central and isolated red galaxies in GAMA exhibit long \tauQ\ relative to \shark, suggesting that this is not the mechanism to reduce the tension between both.
A possible improvement could be to extend the effectiveness of the AGN radio-mode feedback to lower halo masses in \shark, as reducing the amount of gas available for cooling into the galaxy would lead to a gradual decrease of star formation (i.e., a long \tauQ).

\subsection{Comparing our \tauQ\ measurements with literature results}\label{subsec:timescales_comp}

What follows now is a series of comparisons with a representative collection of literature results, both from observations and simulations, which are also shown in Figure \ref{fig:final_fig} as comparison to our results.
Since the calculation of \tauQ\ could have a strong effect on the measured values, we include an overview of how they were derived.
We remark that this means that good quantitative (or even qualitative) agreement should not be expected to be a natural result, as disagreements may well stem from methodological (or even conceptual) differences.
Also, little discussion is presented on the possible lookback time dependence of \tauQ\ in the literature, so we will omit that aspect.

Starting with results from observations, \citet{schawinski2014} inferred colour transition timescales combining colours and stellar masses of galaxies from the Sloan Digital Sky Survey \citep[SDSS;][]{york2000} with the morphological classification from GalaxyZoo \citep{lintott2008}.
For this, they first divided their galaxy sample between "blue", "green" and "red" in the colour-stellar mass plane, with the limits being straight lines chosen by visual inspection (see Figure \ref{fig:PRlit_tLBR_Mstar} and Table \ref{tab:lit}).
They generated simple colour histories with a exponentially-decaying SFH with different $e$-folding timescales, and then compared the colour evolution from these SFHs to the colour-colour distribution of both early- and late-type galaxies.
From these comparisons they inferred a fast transition for early-type galaxies (\tauQ$\lesssim250$ Myr), with late-type galaxies transitioning slowly (\tauQ$\gtrsim1$ Gyr).
Since only $\sim20\%$ of the GAMA sample shown in Figure \ref{fig:final_fig} correspond to elliptical galaxies (according to the \citet{driver2022} morphological classification), the expectation would be for our sample to better match the timescales that \citet{schawinski2014} found for late-type galaxies, which is indeed the case.
In contrast, \shark/\sharkfit\ show \tauQ\ in agreement with \citet{schawinski2014} only for galaxies above $\sim$\mstar{10.5}, with galaxies below this mass exhibiting a median \tauQ\ neither consistent with early-type nor late-type galaxies.
While \citet{schawinski2014} find a dependence of \tauQ\ on halo mass for late-type galaxies, they do not find the same for the early-types that dominate the red population, which agrees with the similarity we find between low- and high-multiplicity groups in GAMA.

\citet{smethurst2015} also used GALEX/SDSS photometry plus GalaxyZoo morphological classification \citep[though the more recent GalaxyZoo2 release,][]{willett2013} to infer quenching timescales.
They also classify galaxies as blue, green or red, but use instead the definition from \citet{baldry2004}, where everything $<1\sigma$ from the local minimum in colour-magnitude as green.
Furthermore, they also adopt a similar approach of generating sample colour evolution tracks from exponentially-decaying SFHs, though they use a Bayesian approach to find the timescale and quenching onset that best matches every galaxy in their sample.
For bulge- and disc-dominated galaxies they find median timescales of $\sim1$ and $\sim2$ Gyr respectively, which is in good agreement with our results from GAMA.
Their figures 8 and 11 suggest that these values may be strongly driven by galaxies quenching early in the Universe (\tLBR$>6$ Gyr), which we do not explore in this work due to biases in the recovery.
It is interesting to note that they seem to find a strong evolution toward longer timescales at more recent times (see their figures 8 through 11), though it is not clear if this is just a selection effect.
Regardless, this qualitatively agrees with our findings in GAMA, where galaxies that become red more recently do it on longer timescales than those becoming red earlier.

\citet{rowlands2018} used the strength of the 4000 \AA\ break and the excess Balmer absorption from GAMA and the VIMOS Public Extragalactic Redshift Survey (VIPERS).
They first divided galaxies as either post-starburst or not based on being above or below a Balmer absorption limit, and then further divided the non post-starburst between blue, green and red based on two ad hoc 4000 \AA\ break values.
They then measured the number density evolution of these classifications at a wide range of redshifts ($0.05<z<1.0$) to infer transition timescales at two partly overlapping stellar mass ranges ($>$\mstar{10.6} and $>$\mstar{11}).
They found transition timescale for green valley galaxies of $\sim2.6$/$6.6$ Gyr for their mid/high-mass selection, independent of lookback time.
This positive trend for timescales with stellar mass is opposite to our findings in GAMA, but it is likely driven by those being a different type of timescales, with the values themselves being significantly higher than seen in either GAMA or \shark.
This is likely because \citet{rowlands2018} measure the timescale for $z\sim0.7$ green valley galaxies would join the $z=0$ red population, whereas we measure the timescale over which observed red galaxies became red, which \citet{schawinski2014} shows are dominated by different morphological types.

\citet{bremer2018} measured the fraction of galaxies in the green valley as a function of environment from GAMA, defined in colour-stellar mass space (see Figure \ref{fig:PRlit_tLBR_Mstar} and Table \ref{tab:lit}), and combined it with the stellar ages of \citet{taylor2011} to infer \tauQ\footnote{A similar method was also presented in \citet{phillipps2019}, but instead they used the $e$-folding time from the \citet{taylor2011} fits to infer how long will current green galaxies take to become red.
    This is the reason why we do not discuss their results, to avoid comparisons between our \textit{reconstructed} evolution to their \textit{predicted} evolution.}
.
They limited their analysis to galaxies with stellar masses in the $10^{10.25}$--\mstar{10.75} range, $0.1<z<0.2$ and $r$-band axial ratio $b/a>0.5$.
They found a \tauQ\ of  $\sim$1--2 Gyr, in good agreement with our measurement of \tauQ\ for GAMA.
Also similar to our results from GAMA, they found no evidence for environmental effect from the near-constant fraction of green valley galaxies as a function of group multiplicity\footnote{They do find evidence that galaxies in high density environment have shorter lifespans as part of the blue population than in less dense environment, suggesting that a richer environment will trigger an earlier transition to red.}, but our results from \shark\ show that central/satellite confusion can strongly diminish any environmental signature present in \tauQ.

We now focus on a comparison with literature results presented for galaxy formation simulations.
\citet{trayford2016} used galaxies with stellar masses of $10^{10}$--\mstar{11} from the EAGLE simulation \citep{schaye2015}, and classifying them as blue, green or red by ad hoc colour selections, defined as a function of both stellar mass and redshift (see Figure \ref{fig:PRlit_tLBR_Mstar} and Table \ref{tab:lit}).
From these, they selected $z=0$ red galaxies and measured the timescale over which they transition from blue to red.
They found a median \tauQ\ of $\sim2$ Gyr, though with a distribution strongly skewed towards shorter timescales, with a peak closer to $\sim1.5$ Gyr.
While they do not find a strong difference in the median \tauQ\ for centrals and satellites (order of a few hundred Myr), the distributions shown in their figure 10 indicate that centrals are less skewed to short timescales.
\shark\ does display the same trend, though we find shorter \tauQ\ (factor of $\sim3$).
Their results are in agreement with our results from GAMA, but it is not clear if this holds for earlier times, as they do not explore the time evolution of \tauQ.
They do not find evidence of a strong dependence with stellar mass, which seems in better agreement with \shark\ than GAMA.
Finally, we find in \shark\ the same results that they do with regards to satellites: the majority become red when becoming a satellite.

\citet{nelson2018a} employed a similar method to ours to measure \tauQ\ from the IllustrisTNG simulation \citep{pillepich2018}, first characterising the colour population with two Gaussian components, with parameters as function of stellar mass and redshift.
They then defined the limits for each population, set at 1$\sigma$ from the mean of each population, which is the most significant difference with our probability-based approach (see Figure \ref{fig:PRlit_tLBR_Mstar} and Table \ref{tab:lit}).
Like \citet{trayford2016}, they also found an asymmetrical \tauQ\ distribution, skewed to shorter values, finding similar median and peak values ($\sim2$ and $\sim1.6$ Gyr, respectively).
They found a dependence with stellar mass, with \tauQ\ peaking for galaxies of $\sim$\mstar{10}.
While the range of median \tauQ\ values they measure coincides with that from GAMA, we find a different trend with stellar mass, with the best agreement being for galaxies $\gtrsim$\mstar{10.5}.
They also found a weak trend for centrals below \mstar{10} to take longer to become red than satellites.

Also using galaxies from the EAGLE simulation, \citet{wright2019} used a classification close to that used by \citet{nelson2018a}, finding \tauQ\ to be in the $\sim$2--4 Gyr range, depending on both stellar mass and environment.
They found different \tauQ\ for centrals and satellites for galaxies below $\sim$\mstar{10.5} ($\sim4$ and $\sim2$ Gyr respectively), with timescales showing a inverted U-shape and both centrals and satellites peaking at $\sim$\mstar{9.7}.
For larger stellar masses they found all galaxies to have similar \tauQ\ ($\sim2$ Gyr).
Their \tauQ\ measurements are in strong agreement with those of \citet{nelson2018a}, despite using different thresholds to measure \tauQ\ (see Table \ref{tab:lit}), which suggests that the stellar mass trend of \tauQ\ found in both works may be a consequence of how they define the limits of the blue and red populations.

\section{Conclusions}\label{sec:summary}

In this work, we have used the characterisation of the colour evolution of the blue and red galaxy populations we presented in \citetalias{bravo2022}, to calculate upper limits for \tauQ\ on which red galaxies transitioned from being blue to red.
For this, we first calculated the probability of all galaxies in our three samples (GAMA, \shark\ and \sharkfit) to belong to the red population, then used the distribution of this probability to define the values between which we will measure \tauQ.
Accounting for selection biases, we find evidence that \tauQ\ evolves with time only in GAMA, with \tauQ\ increasing from $\sim1$ to $\sim3$ Gyr in a time span of $\sim4$ Gyr (in \shark/\sharkfit\ \tauQ\ remains stable at $\lesssim1$ Gyr).
Our observations and simulations do not agree on whether there is a stellar mass dependence on the lookback time when they became red, with the former in agreement with a large body of existing literature that current high-mass galaxies became red before low-mass galaxies (i.e., downsizing), while the latter fails to reproduce such trend.
We find a difference between centrals and satellites in GAMA only for $M_\star\lesssim$\mstar{10}, with satellites showing \tauQ\ $\sim0.4$ Gyr shorter than centrals.
The results from \shark\ suggest the possibility of a larger difference being hidden by observational central/satellite classification confusion.
Finally, we find that assuming an instantaneous halo gas stripping in \shark\ is the likely driver for the shorter-than-observed \tauQ\ for satellites.

\section*{Acknowledgements}

We thank Chris Power and Pascal Elahi for their role in completing the SURFS $N$-body DM-only simulations suite, Rodrigo Tobar for his contributions to \shark, Andrea Cattaneo and Benjamin Johnson for the comments and feedback provided to the doctoral thesis on which this work is based, Ruby Wright for providing the data from \citet{wright2019} for Figure \ref{fig:final_fig}.

MB acknowledges the support of the University of Western Australia through a Scholarship for International Research Fees and Ad Hoc Postgraduate Scholarship, and the funding by McMaster University through the William and Caroline Herschel Fellowship.
LJMD and ASGR acknowledge support from the Australian Research Councils Future Fellowship scheme (FT200100055 and FT200100375, respectively).
CdPL is funded by the ARC Centre of Excellence for All Sky Astrophysics in 3 Dimensions (ASTRO 3D), through project number CE170100013.
CdPL also thanks the MERAC Foundation for a Postdoctoral Research Award.
SB acknowledges support by the Australian Research Council’s funding scheme DP180103740.
JET is supported by the Australian Government Research Training Program (RTP) Scholarship.

This work was supported by resources provided by the Pawsey Supercomputing Centre with funding from the Australian Government and the Government of Western Australia.
We gratefully acknowledge DUG Technology for their support and HPC services.

GAMA is a joint European-Australasian project based around a spectroscopic campaign using the Anglo-Australian Telescope.
The GAMA input catalogue is based on data taken from the Sloan Digital Sky Survey and the UKIRT Infrared Deep Sky Survey.
Complementary imaging of the GAMA regions is being obtained by a number of independent survey programmes including GALEX MIS, VST KiDS, VISTA VIKING, WISE, Herschel-ATLAS, GMRT and ASKAP providing UV to radio coverage.
GAMA is funded by the STFC (UK), the ARC (Australia), the AAO, and the participating institutions.
The GAMA website is \url{http://www.gama-survey.org/}.
Based on observations made with ESO Telescopes at the La Silla Paranal Observatory under programme ID 179.A-2004.
Based on observations made with ESO Telescopes at the La Silla Paranal Observatory under programme ID 177.A-3016.

The analysis on this work was performed using the programming languages \textsc{Python} v3.10 (\url{https://www.python.org}), with the open source packages \textsc{matplotlib} v3.7 \citep{matplotlib}, \textsc{NumPy} v1.24 \citep{numpy}, \textsc{pandas} v1.5 \citep{pandas}, \textsc{SciCM} v1.0 (\url{https://github.com/MBravoS/scicm}), \textsc{SciPy} v1.10 \citep{scipy}, and \textsc{splotch} v0.6 (\url{https://github.com/MBravoS/splotch}), in addition of the software previously described.

\section*{Data Availability}

The \PR\ tracks and \tauQ\ catalogues generated for this work will be shared on reasonable request to the corresponding author.
For all other data, see the Data Availability statement in \citetalias{bravo2022}.

\bibliographystyle{mnras}
\bibliography{papers} 

\appendix

\section{Recovery of \tauQ\ from \shark\ with \prospect}\label{app:recovery}

\begin{figure}
    \centering
    \includegraphics[width=\linewidth]{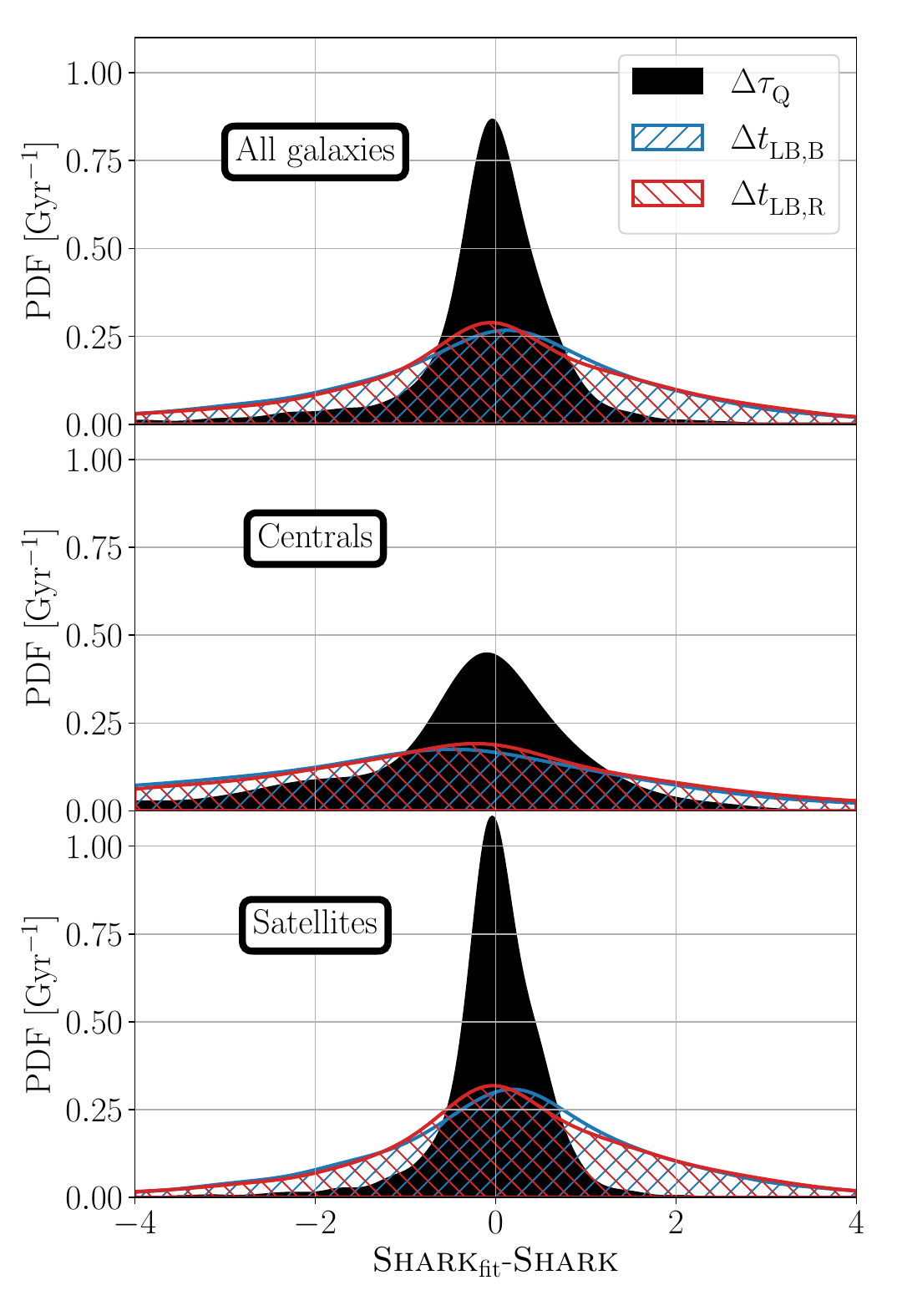}
    \caption{Recovery of \tLBB\ (in hatched blue), \tLBR\ (hatched red), and \tauQ\ (solid black) from \shark\ red galaxies with \prospect.
    To avoid visualisation artefacts due to the discreteness of all three values shown ($\Delta$\tLBB, $\Delta$\tLBR, and $\Delta$\tauQ), the PDFs shown have been constructed using the \texttt{gaussian\_kde} Gaussian Kernel Density Estimator (KDE) function from \textsc{scipy}.
    \comment{maybe add a few statistics to each panel?}}
    \label{fig:SHARK_recovery_all}
\end{figure}

Figure \ref{fig:SHARK_recovery_all} shows the recovery of \tLBB, \tLBR, and \tauQ\ of \shark\ galaxies using \prospect.
In general, we find small median biases in the recovery of \tauQ\ ($\lesssim0.03$ Gyr) but we find a large scatter in the recovery (16$^\mathrm{th}$--84$^\mathrm{th}$ percentile range of $\sim1.1$ Gyr), indicating that the population as a whole is reasonable recovered but not individual galaxies.
\tauQ\ is better recovered than either \tLBB\ or \tLBR, e.g., we better recover \textit{how fast} galaxies become red rather than \textit{when} they leave (enter) the blue (red) population.

\begin{figure}
    \centering
    \includegraphics[width=\linewidth]{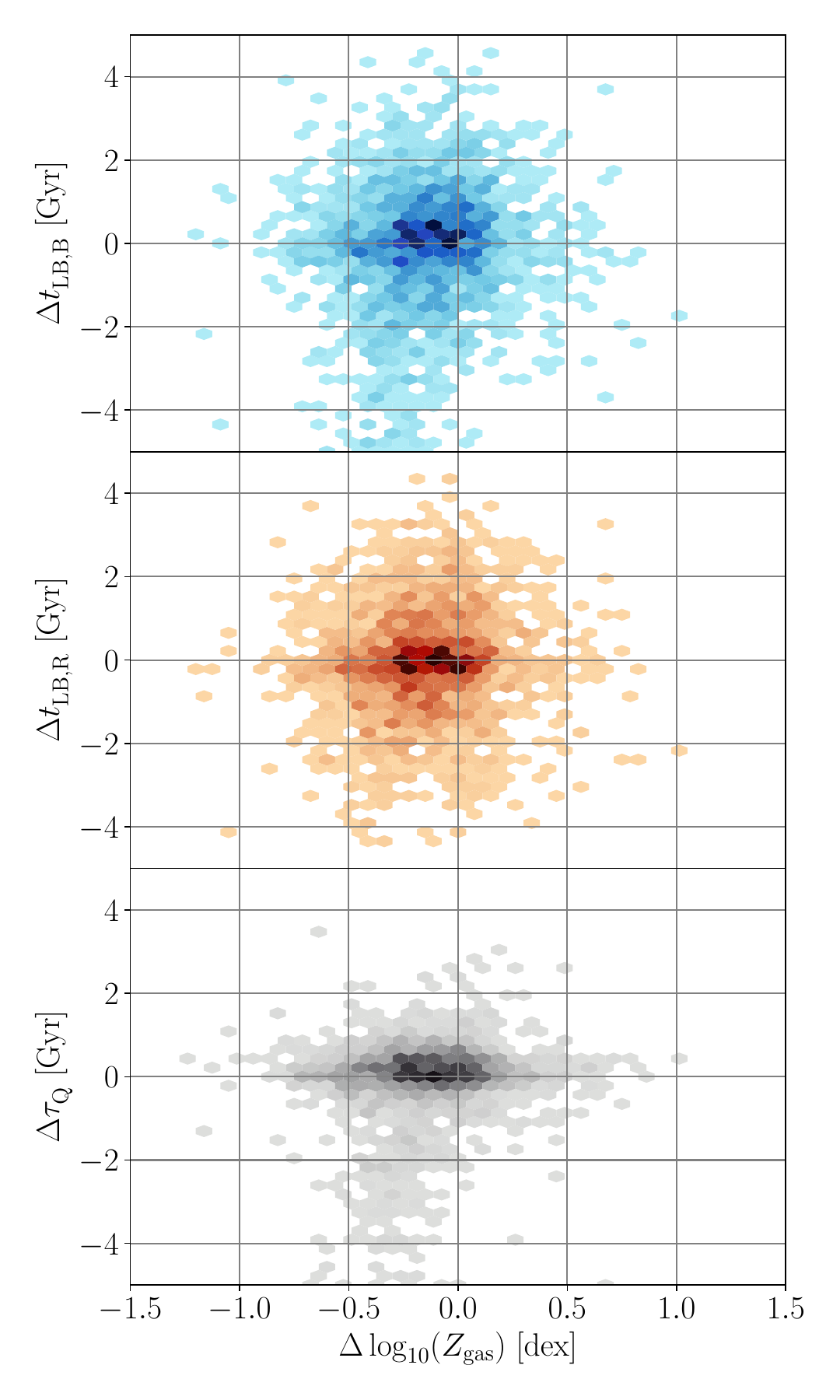}
    \caption{Relation between the recovery of the gas-phase metallicity, $Z_\mathrm{gas}$, to the recovery of \tLBB\ (top panel), \tLBR\ (middle), and \tauQ\ (bottom) with \prospect
    The colour of the bins indicate the number of galaxies with a linear scaling.
    For $\Delta$\tLBB\ and $\Delta$\tauQ\ a few ($\lesssim1.5\%$) lie below their respective $y$-axis lower limits.}
    \label{fig:SHARK_Delta_Z}
\end{figure}

\begin{figure*}
    \centering
    \includegraphics[width=\linewidth]{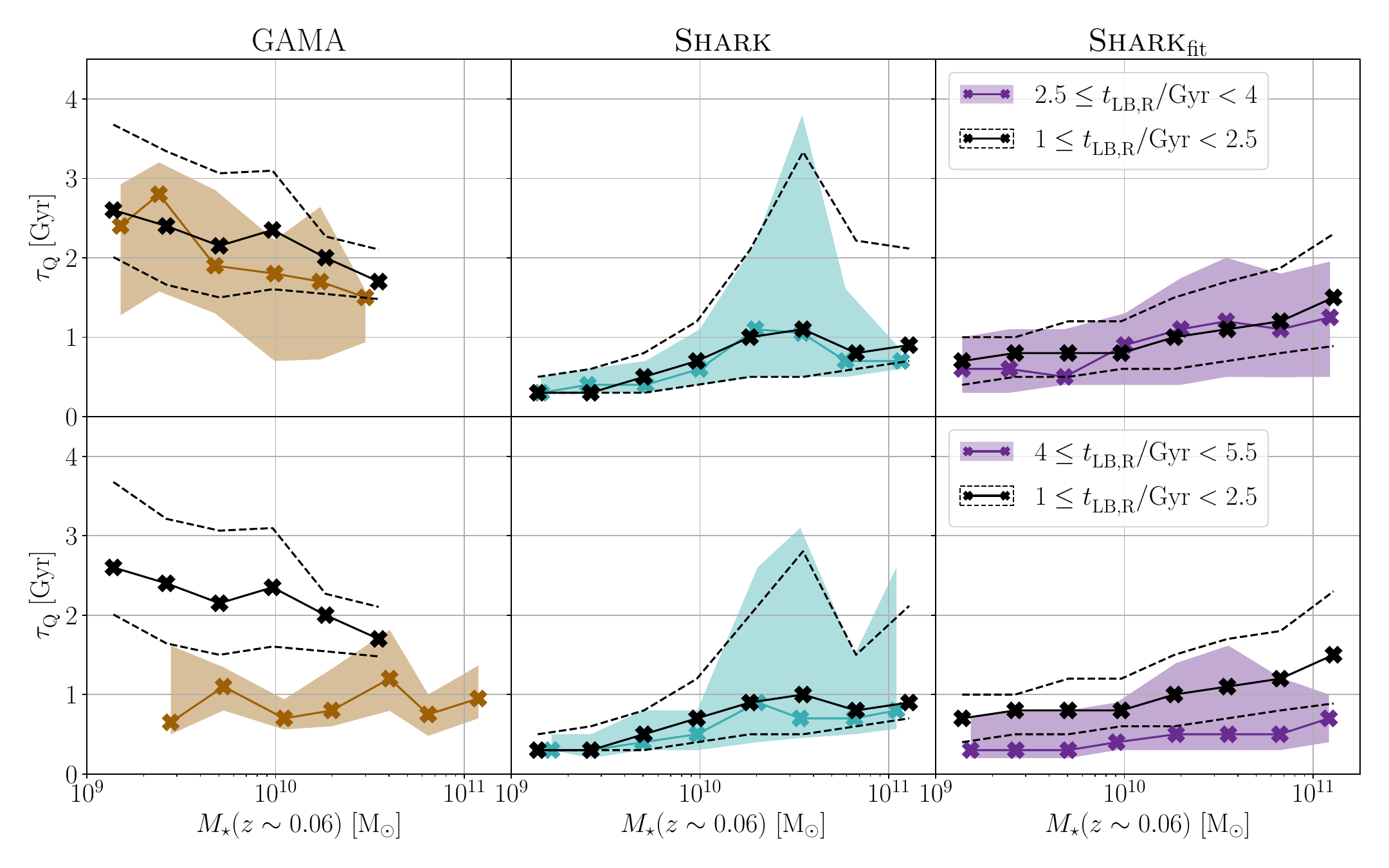}
    \caption{Comparison between the \tauQ\ measured at two different lookback time bins with a comparable \tLBR--\tauQ\ selection.
    The top row compares the \tauQ\ distribution between the $1\leq$\tLBR$/\mathrm{Gyr}<2.5$ and $2.5\leq$\tLBR$/\mathrm{Gyr}<4$ bins, the bottom row between $1\leq$\tLBR$/\mathrm{Gyr}<2.5$ and $4\leq$\tLBR$/\mathrm{Gyr}<5.5$.
    The solid lines indicate the \tauQ\ running median, and the dashed lines and shaded areas the 16-84$\mathrm{th}$ percentiles, with those in colour being measured at the respective \tLBR\ bins and those in black from the lowest \tLBR\ bin.
    Each column shows the results for a different sample, left to right: GAMA, \shark, and \sharkfit.}
    \label{fig:all_Mstar_tauB2R_evol}
\end{figure*}

We find that $\Delta$\tauQ\ trends with other properties, like stellar mass or infall category (see Section \ref{subsec:shark_infall}), are almost completely accounted for by the central/satellite classification, with centrals showing a worse recovery than satellites.
E.g., we find that \tauQ\ recovery worsens with increasing mass, and that category \ref{item:SharkAI} satellites are better recovered than those in category \ref{item:SharkBI}, but those are a consequence of the dominating type of galaxies as a function of stellar mass and that category \ref{item:SharkAI}/\ref{item:SharkBI} galaxies quenched as satellites/centrals.
The one exception to this are $\sim$\mstar{10.5} centrals, which are the main driver for the skew towards under-estimated \tauQ\ values.
The difference between centrals and satellites is likely a consequence of the SFH model we use in \prospect, a skewed-Gaussian SFH, being better suited to model the quenching of the latter.
This should not be necessarily understood as rejuvenation being a key factor, as few red galaxies undergo a rejuvenation episode ($\lesssim3$\%, see appendix A of \citetalias{bravo2022}), but rather that limited gas replenishment can extend the time quenching in a manner that is not well-captured by a skewed Gaussian.

We also tested a different choice for \tLBB\ than the one outlined in Section \ref{sec:timescales}, using the first time the galaxy left the blue population instead of the last.
This was motivated by the assumption that the skewed Normal SFH we adopt in \prospect\ would more likely recover an smooth proxy of the true \shark\ SFHs, with the longer \tauQ\ definition leading to a better match with \sharkfit.
We found the opposite to be true, with this longer \tauQ\ definition leading to a strong bias in the recovery of both \tLBB\ and \tauQ\  ($\sim$1--2 Gyr, depending on stellar mass), with the shorter definition leading to a less biased recovery (see Appendix \ref{app:recovery}).
This finding suggests that our choice of SFH is robust against multiple episodes of significant star formation and provides a sensible recovery of the last quenching episode\footnote{Which is not to say that the overall evolution of the galaxy would be well-recovered, which depends on how well the overall evolution of the galaxy can be approximated with the models adopted in \prospect.}, which we have verified by examining individual SED fits against the true evolution in \shark.

\shark\ galaxies that remain for an extended period in the transitional region have proven to a more significant challenge, as can be seen in Figures \ref{fig:popfrac_censat} and \ref{fig:all_Mstar_tauB2R}.
The bottom panel in Figure \ref{fig:SHARK_Delta_Z} shows that, while there is no particular correlation between $\Delta$\tauQ\ and the gas-phase metallicity ($Z_\mathrm{gas}$) for most galaxies, galaxies with $\Delta$\tauQ$\lesssim-2$ Gyr show a systematic bias in the recovery of $Z_\mathrm{gas}$.
From the upper and middle panels in Figure \ref{fig:SHARK_Delta_Z} clearly indicate that the error in \tauQ\ is driven by errors in \tLBB, indicating that these galaxies are being recovered as star-forming for a longer period than in \shark.

This finding is in line with the difference between the true and recovered SFHs shown in figure A.1 in \citetalias{bravo2022}.
These are $\gtrsim$\mstar{10.5}, \tauQ$\gtrsim2$ Gyr galaxies (see Figure \ref{fig:all_Mstar_tauB2R}), which correspond to the galaxies we highlighted in appendix A.3 of \citetalias{bravo2022} as particularly challenging to fit.
We found that this stems from dust properties of these bulge-dominated galaxies, with steep dust attenuation slopes (relative to discs) that proved challenging to recover with our current modelling choices in \prospect, leading to biases also in their recovered SFHs and $Z$Hs.
This is unlikely to significantly affect our measurements and conclusions, since the \sharkfit\ results indicate that this only leads to an underestimation in the dispersion of \tauQ.


\section{Corroboration of the time evolution of \tauQ, or lack of thereof}\label{app:timescale_evol}

To explore if any of our samples display a time-dependent \tauQ\ distribution, when comparing two lookback time bins set a limit to the maximum \tauQ\ included in the comparison.
This is to remove the possible bias due to the larger span of \tauQ\ values that we can measure at the lower lookback time bin.
I.e., when comparing the $1\leq$\tLBR$/\mathrm{Gyr}<2.5$ and $2.5\leq$\tLBR$/\mathrm{Gyr}<4$ bins, we set the upper \tauQ\ limit for both bins at 6 Gyr, as that is the largest \tauQ\ we can measure at a looback time of 4 Gyr given our chosen starting point of 10 Gyr (Section \ref{sec:GSP}).
This process ensures an equal \tauQ\ completeness for the two lookback time bins being compared, enabling us to study whether the \tauQ\ distribution evolves with cosmic time or not.

Figure \ref{fig:all_Mstar_tauB2R_evol} shows the measured \tauQ\ as a function of stellar mass at two different \tLBR\ bins ($2.5\leq$\tLBR$/\mathrm{Gyr}<4$ and $4\leq$\tLBR$/\mathrm{Gyr}<5.5$) and compares them to the lowest \tLBR\ bin ($1\leq$\tLBR$/\mathrm{Gyr}<2.5$).
\shark\ shows a strong consistency when comparing similarly-selected samples at different lookback times, indicating that \tauQ\ does not depend on lookback time for this sample.
In contrast, GAMA exhibits a strong evolution.
Galaxies with $M_\star<\sim$\mstar{10.5} that became red at a lookback time of 4--5.5 Gyr have \tauQ\ values that are a factor of $\sim2$ shorter than those of similarly-selected galaxies that transitioned in the 1--2.5 Gyr range, with galaxies above that stellar mass showing a smaller evolution in \tauQ.
A similar decrease by a factor of $\sim2$ is evident in \sharkfit, though without the stellar mass dependence seen in GAMA.
While this suggests that this evolution is at least partially due to our modelling choices in \prospect, it is not obvious that this can fully explain it, as GAMA and \sharkfit\ display different mass dependencies and measured timescales.

\bsp	
\label{lastpage}
\end{document}